\documentclass[journal]{IEEEtran}

\usepackage{amsmath, amssymb, amsthm}  
\allowdisplaybreaks

\usepackage{algorithm}      
\usepackage{algpseudocode}
\usepackage{graphicx}
\usepackage{subfigure}
\usepackage{tabularx} 
\usepackage{caption}  
\usepackage{subcaption}  
\usepackage{booktabs}  
\usepackage{multirow}  
\usepackage{url}  
\usepackage{cite}  
\usepackage{color}  
\usepackage{xcolor}  
\usepackage{hyperref}  
\usepackage{siunitx}  
\usepackage{mathrsfs}  
\usepackage{mathtools}  
\usepackage{txfonts}  
\usepackage{fontawesome5}  
\usepackage{epstopdf}  
\usepackage{enumitem}  
\usepackage{pgfplots}  
\usepackage{verbatim}  
\usepackage{placeins}  
\usepackage{pdflscape}  
\usepackage{balance}  
\usepackage{lipsum}  
\usepackage{upgreek}  
\usepackage{amsfonts} 
\usepackage{bm}  

\makeatletter
\newenvironment{breakablealgorithm}
{
		\begin{center}
			\refstepcounter{algorithm}
			\hrule height.8pt depth0pt \kern2pt
			\renewcommand{\caption}[2][\relax]{
				{\raggedright\textbf{\ALG@name~\thealgorithm} ##2\par}%
				\ifx\relax##1\relax 
				\addcontentsline{loa}{algorithm}{\protect\numberline{\thealgorithm}##2}%
				\else 
				\addcontentsline{loa}{algorithm}{\protect\numberline{\thealgorithm}##1}%
				\fi
				\kern2pt\hrule\kern2pt
			}
		}{
		\kern2pt\hrule\relax
	\end{center}
}
\makeatother

\title{Where to Perform Channel Measurements for CKM Construction: A Random Field Theory Analysis }

\vspace{-2ex}

\author{
	\IEEEauthorblockN{Yuxuan Song, Yong Zeng,~\IEEEmembership{Fellow,~IEEE}, Xiping Wu,~\IEEEmembership{Senior Member,~IEEE}, Guangchi Zhang, \\Cheng-Xiang Wang,~\IEEEmembership{Fellow,~IEEE}, and Rui Zhang,~\IEEEmembership{Fellow,~IEEE}}
	\thanks{

		Yuxuan Song, Yong Zeng, Xiping Wu and Cheng-Xiang Wang are with the National Mobile Communications Research Laboratory, Southeast University, Nanjing 211189, China. Yong Zeng, Xiping Wu and Cheng-Xiang Wang are also with the Purple Mountain Laboratories, Nanjing 211111, China (e-mail: \{220240906, yong\_zeng, xiping.wu, chxwang\}@seu.edu.cn). (Corresponding author: Yong Zeng.)

		Guangchi Zhang is with the School of Information Engineering, Guangdong University of Technology, Guangzhou 510006, China (e-mail: gczhang@gdut.edu.cn).

		Rui Zhang is with the Department of Electrical and Computer Engineering, National University of Singapore, Singapore 117583 (e-mail: elezhang@nus.edu.sg).
}}

\begin{document}
	\maketitle

	\vspace{-0.2in}

	\begin{abstract}
		Channel knowledge map (CKM) is regarded as a promising technology for future sixth-generation (6G) networks, facilitating environmental-aware wireless communication, sensing, and localization. Research works on CKM construction can be classified as model-based methods and data-based approaches. Specifically, data-based CKM construction exploits the fundamental principle of spatial correlation to complete CKM based on limited measurement data, leading to the question of ``where to perform channel measurements''. In this paper, we study the spatial measurement strategy for efficient data-based CKM construction, and consider a specific type of CKM named channel gain map (CGM). The general objective is to select a subset of locations for channel measurements, so as to minimize the average mean-squared-error (AMSE) of the global CGM construction. In order to reduce the infinite measurement locations to a finite set, we discretize the underlying physical space into a finite number of cubic grid points, and formulate a combinatorial optimization problem to select measurement locations from them. In order to solve the proposed problem, we employ two representative algorithms, namely the greedy algorithm and the simulated annealing (SA), and discuss their respective advantages. To overcome the accuracy-complexity trade-off of traditional uniform discretization, we develop an adaptive discretization strategy from the viewpoint of Gaussian random field theory to minimize the information loss from the original continuous field to its approximated discrete representation in the mean-squared sense. Compared to uniform discretization, the proposed adaptive discretization strategy achieves a significant performance gain in terms of AMSE-reduction, establishing the theoretical framework of spatial measurement and providing practical guidance for implementation.
		
		\begin{IEEEkeywords}
			Channel knowledge map, Kriging interpolation, Gaussian field theory, combinatorial optimization.
		\end{IEEEkeywords}

	\end{abstract}

	\section{Introduction}
	
	Channel knowledge map (CKM) \cite{Zeng2021Mag, zeng2024tutorial, 6G} is a key enabler for future sixth-generation (6G) wireless networks. The fundamental idea behind CKM is to establish the mapping relationship from physical/virtual locations to a high-dimensional channel space, where key channel parameters, such as amplitude, delay and angle-of-arrival (AoA), are incorporated into a unified framework for facilitating environmental-aware communication, sensing, and localization.
	
	In particular, channel gain map (CGM) is an important type of CKM that aims to learn the channel gain between a serving base station (BS) and any location within the cell. Research works on CGM construction can be divided into model-based methods \cite{ScatterCKM,Semantics,ModelAccess,ModelTWC,xu2024much,Continuous4,Virtual2} and data-based approaches\cite{DistributedKriging,2024Yang,SongTCOM,CKMDiff,Continuous3,SparseMI}. Model-based methods typically formulate a closed-form propagation model characterized by undetermined parameters. Large-scale propagation factors, such as path-loss and shadowing, are usually separated from random small-scale fading, and the parameters are fitted based on measurement data. The advantages of model-based methods lie in their theoretical rigor and robustness to measurement noise, among others. Nevertheless, due to the complex nature of wireless propagation, it is difficult to formulate a universal model that fully captures the spatial inhomogeneity \cite{ModelTWC,localizationTWC} and temporal variability\cite{CKMUpdate,AoIUpdate} of location-dependent channel gain. Therefore, the performance of model-based methods is prone to model mismatch, and the fine-grained local features of channel variations tend to be smoothed out. Data-based approaches, on the other hand, do not make explicit assumptions about the underlying channel model, but instead make use of the inherent spatial correlation between measured and unmeasured locations. The interpolation for channel gains in unmeasured locations is performed in a local neighborhood characterized by the local spatial correlation distance, thus precluding the propagation of global errors induced by local anomalous data. Compared to model-based methods, data-based strategies are generally more flexible but more dependent on high-quality measurement data to learn the specific structure of CGM.

	Therefore, one of the most important tasks for data-based CGM construction is to determine the optimal measurement locations from all candidate measurement locations, which falls into the category of combinatorial optimization. At its core, the ground-truth CGM is continuous by nature, a characteristic resulting from the intrinsic continuity of the two-dimensional (2D) or three-dimensional (3D) physical space. In line with this principle, an increasing number of channel modeling studies have placed an emphasis on continuous 3D spatial modeling \cite{Continuous1,Continuous2,6G2,6G3}, which aims to accurately capture the continuous variation characteristics of wireless channels across 3D physical space. Nevertheless, the continuous modeling introduces an infinite number of candidate measurement locations. In order for a combinatorial optimization problem with an infinite feasible set to be solvable, the objective function and constraints need to possess several favorable mathematical properties (e.g., continuity, differentiability, monotonicity) so that the infinite feasible set can be reduced to a finite one \cite{OptimizationAlgp}. Unfortunately, such conditions generally do not hold for CGM. To be concrete, due to the line-of-sight (LoS) obstruction by sharp-edged environmental scatterers such as high-rise buildings, there could be abrupt channel gain transitions in the boundaries between LoS and non-line-of-sight (NLoS) regions, undermining the smoothness of the channel gain function. Due to the dynamic nature of wireless propagation environments, the small-scale fading in a spatial region tends to de-correlate rapidly in both time and space, contributing to non-monotonic channel gain variations that are generally hard to predict. Therefore, in order to find the optimal measurement locations, it is reasonable to discretize the underlying continuous space in the first place.

	Most previous works on CGM construction adopt the method of uniform-granularity discretization, where the continuous 2D/3D space is discretized into equal-sized square/cubic grid points. Nevertheless, such a strategy inevitably results in a tradeoff between CGM accuracy and CGM measurement/storage complexity. On one hand, a large granularity alleviates the stress of CGM construction/completion but at the risk of missing important channel knowledge in spatially fast-varying regions. For example, in ground-to-air communication, due to frequent LoS obstruction and the existence of beampattern nulls, well-covered aerial regions and potential coverage holes could exhibit an alternating spatial distribution. A large discretization granularity may fail to identify hidden coverage holes\cite{zeng2021simultaneous,xu2019cellular,LuTWC}, potentially resulting in catastrophic communication outages. On the other hand, a small granularity prioritizes CGM accuracy but introduces extra storage burden, and raises the risk of combinatorial search complexity explosion when determining the optimal measurement locations.

	According to the basic spatial field theory \cite{RandomField}, many location-dependent physical phenomena, such as temperature and channel gain, evolve with time and can be modeled as the sum of a deterministic trend function and a zero-mean spatial random field. The channel gain at a specific location fluctuates around its mean value, and the variance of fluctuation depends on the stationarity of the underlying spatial field. Such a stochastic viewpoint acknowledges the dynamic nature of complex physical phenomena and replaces traditional deterministic channel gain mapping with location-conditional probability density function (PDF), capturing the variability of location-dependent channel gain in the statistical sense. Specifically, the covariance function, as the second-order statistical property, can be used to determine the spatial correlation distance of a specific spatial subregion, quantifying the level of channel gain variation. Based on this spatial information, the limited grid resources for spatial measurements can be adaptively allocated among different subregions by adjusting the corresponding discretization granularities. 
	
    However, the introduction of the random field theory and a design of adaptive discretization only capture half of the picture. In order to determine the optimal measurement locations, a suitable mathematical metric is needed to predict the mean-squared-error (MSE) at an unmeasured location based on a set of potential measurement locations. Kriging, as a class of linear unbiased minimum-mean-squared-error (MMSE) estimator \cite{Kriging,measureCKM,ReKriging}, is a useful tool to provide such prediction. Specifically, the unbiasedness of Kriging ensures that the MSE for estimating the channel gain at a specific unmeasured location is equivalent to the closed-form Kriging variance in the statistical sense. Therefore, the average Kriging variance at all unmeasured locations can serve as a tractable optimization objective to approximate the average MSE (AMSE) of the global CGM construction, allowing for the global identification of measurement locations that contribute to high AMSE-reduction. In this paper, we aim to select a subset of locations for channel measurements from a discrete universal set of possible measurement locations, so as to minimize the AMSE of channel gains estimation at those unmeasured locations based on the measurement data. Our main contributions are summarized as follows:

	\begin{enumerate}[label={\textbullet}]
		\item We formulate an optimization problem to determine the optimal set of locations to perform channel measurements for CGM construction, and use the closed-form Kriging variance to quantify the AMSE. The optimization problem provides a general framework for conducting spatial measurements, and we solve the formulated problem using a greedy algorithm or the simulated annealing (SA).
		\item In order to overcome the accuracy-complexity tradeoff of traditional uniform discretization, we develop an adaptive discretization strategy based on the Gaussian random field theory. Specifically, we propose a K-means-based clustering strategy to partition the physical space into homogeneous local subregions. In order to minimize the global information-loss of spatial discretization in the mean-squared sense, we use the Lagrangian multiplier to determine the discretization granularity of each subregion according to its volume and local correlation distance.	
		\item In order to study the hierarchical relationship between spatial discretization and combinatorial algorithmic optimization, we comprehensively evaluate the performance on AMSE-reduction of different optimization methods. Compared to uniform discretization, the proposed adaptive discretization strategy achieves a twenty-percent performance gain in terms of AMSE-reduction.

	\end{enumerate}
	
	The remainder of this paper is organized as follows. Section II presents the system model, and formulates the general optimization problem. Section III studies the formulated problem for the Kriging-based method. Section IV discusses several elementary measurement scenarios to gain some insight into the optimal measurement locations. Section V explains the proposed algorithms. Section VI discusses the adaptive discretization strategy based on the Gaussian random field theory. Section VII demonstrates the performance superiority of the proposed adaptive discretization through extensive simulation results. In the end, Section VIII concludes the paper. The key symbol definitions are summarized in Table 1.
	
	\begin{table}[htbp]
		\centering
		\caption{Summary of the key symbol definitions}
		
		\newcolumntype{L}{>{\bfseries\centering\arraybackslash}p{0.25\linewidth}}
		\begin{tabularx}{\linewidth}{ 
				L  
				X  
			}
			\hline
			\textbf{Symbol} & \textbf{Definition} \\ \hline
			$\boldsymbol{\mathcal{D}}$ & The universal set obtained by uniformly discretizing the continuous physical space.  \\ \hline
			$\mathcal{D}_r,1\leq r\leq R$ & The universal set in the $r$-th subregion. $\boldsymbol{\mathcal{D}}=\mathcal{D}_1\cup \mathcal{D}_2... \cup \mathcal{D}_R$, and $\mathcal{D}_i\cap \mathcal{D}_j=\emptyset,i\neq j$. \\ \hline 
            $\boldsymbol{\mathcal{U}}\subseteq \boldsymbol{\mathcal{D}}$ & The reduced candidate set in which the measurement locations are selected.  \\ \hline
            $\mathcal{U}_r,1\leq r\leq R$ & The reduced candidate set in the $r$-th subregion. $\boldsymbol{\mathcal{U}}=\mathcal{U}_1\cup \mathcal{U}_2... \cup \mathcal{U}_R$, and $\mathcal{U}_i\cap \mathcal{U}_j=\emptyset,i\neq j$. \\ \hline
            $\boldsymbol{\mathcal{S}}\subseteq \boldsymbol{\mathcal{D}}$ & The set of locations to perform channel measurements to minimize the global AMSE.  \\ \hline
			$ \Gamma(\boldsymbol{x}),\hat \Gamma(\boldsymbol{x})$ & The channel gain at spatial location $\boldsymbol{x}$, and its prediction. \\ \hline
			$\mu(\boldsymbol{x})$ & The deterministic mean value of $\Gamma(\boldsymbol{x})$.  \\ \hline
			$S(\boldsymbol{x})$ & The zero-mean Gaussian random field with location-dependent variance $\sigma^2(\boldsymbol{x})$.  \\ \hline
			$C(\boldsymbol{x},\boldsymbol{x}')$ & The covariance function of the spatial field $\Gamma(\boldsymbol{x})$ in the general form.  \\ \hline
			$L_c^r$ & The spatial correlation distance in the $r$-th subregion. \\ \hline
			$\sigma^2_r$ & The channel gain variance for each location in the $r$-th subregion. \\ \hline
			$C_{r}(\boldsymbol{\tau})$ & The covariance function in the $r$-th subregion. \\ \hline
			$S_r(\boldsymbol{k})$ & The power spectral density as the 3D Fourier transform of $C_{r}(\boldsymbol{\tau})$. \\ \hline
			$\Delta_r$ & The discretization granularity in $\mathcal{D}_r$. \\ \hline
			$D_r$ & The discretization loss for each location in the $r$-th subregion.  \\ \hline
			$n_r$ & Number of grids allocated for the $r$-th subregion, and $|\mathcal{U}_r|=n_r$.  \\ \hline
		\end{tabularx}
	\end{table}
	
	\section{System Model}

	In this paper, we consider a typical 3D physical space $\boldsymbol{\mathcal{D}}$ with one BS. The channel gain at a specific location $\boldsymbol{x} \in \mathbb{R}^3$ from the BS is denoted as $\Gamma(\boldsymbol{x})$. CGM can be mathematically abstracted as a mapping relationship from $\mathbb{R}^3\to \mathbb{R}^+$, where each spatial location $\boldsymbol{x}\in \mathbb{R}^3$ is mapped to a specific channel gain $\Gamma(\boldsymbol{x})\in\mathbb{R}^+$. Specifically, $\Gamma(\boldsymbol{x})$ is modeled as 
	
	\begin{equation}
		\Gamma(\boldsymbol{x}) = \mu(\boldsymbol{x})+S(\boldsymbol{x}) > 0, \boldsymbol{x}\in \boldsymbol{\mathcal{D}} ,  \label{$mu+S$}
	\end{equation} where $\mu(\boldsymbol{x})$ is a deterministic spatial mean function, and $S(\boldsymbol{x})$ is a zero-mean spatial random field \cite{RandomField}. We assume that each $S(\boldsymbol{x})$ is a Gaussian random variable with location-dependent variance, i.e., $S(\boldsymbol{x}) \sim \mathcal{N}\left(0, \sigma^2(\boldsymbol{x}) \right), \boldsymbol{x} \in  \mathbb{R}^3$, and $\Gamma(\boldsymbol{x}) \sim \mathcal{N}\left(\mu(\boldsymbol{x}), \sigma^2(\boldsymbol{x}) \right), \boldsymbol{x} \in  \mathbb{R}^3$. Note that the zero-mean residue term $S(\boldsymbol{x})$ is assumed to be practically bounded so that the channel gain $\Gamma(\boldsymbol{x})$ is always positive. The channel gain covariance between two locations $\boldsymbol{x}_i$ and $\boldsymbol{x}_j$ is 
	
	\begin{equation}
		C_{ij}=\mathbb{E}\left[\left( \Gamma(\boldsymbol{x}_i)-\mu(\boldsymbol{x}_i)\right)\left(\Gamma(\boldsymbol{x}_j)-\mu(\boldsymbol{x}_j)\right)\right]. \label{covariance111}
	\end{equation}

	\begin{figure}[H]	
		\centering	
		\includegraphics[width=9cm,height=5cm]{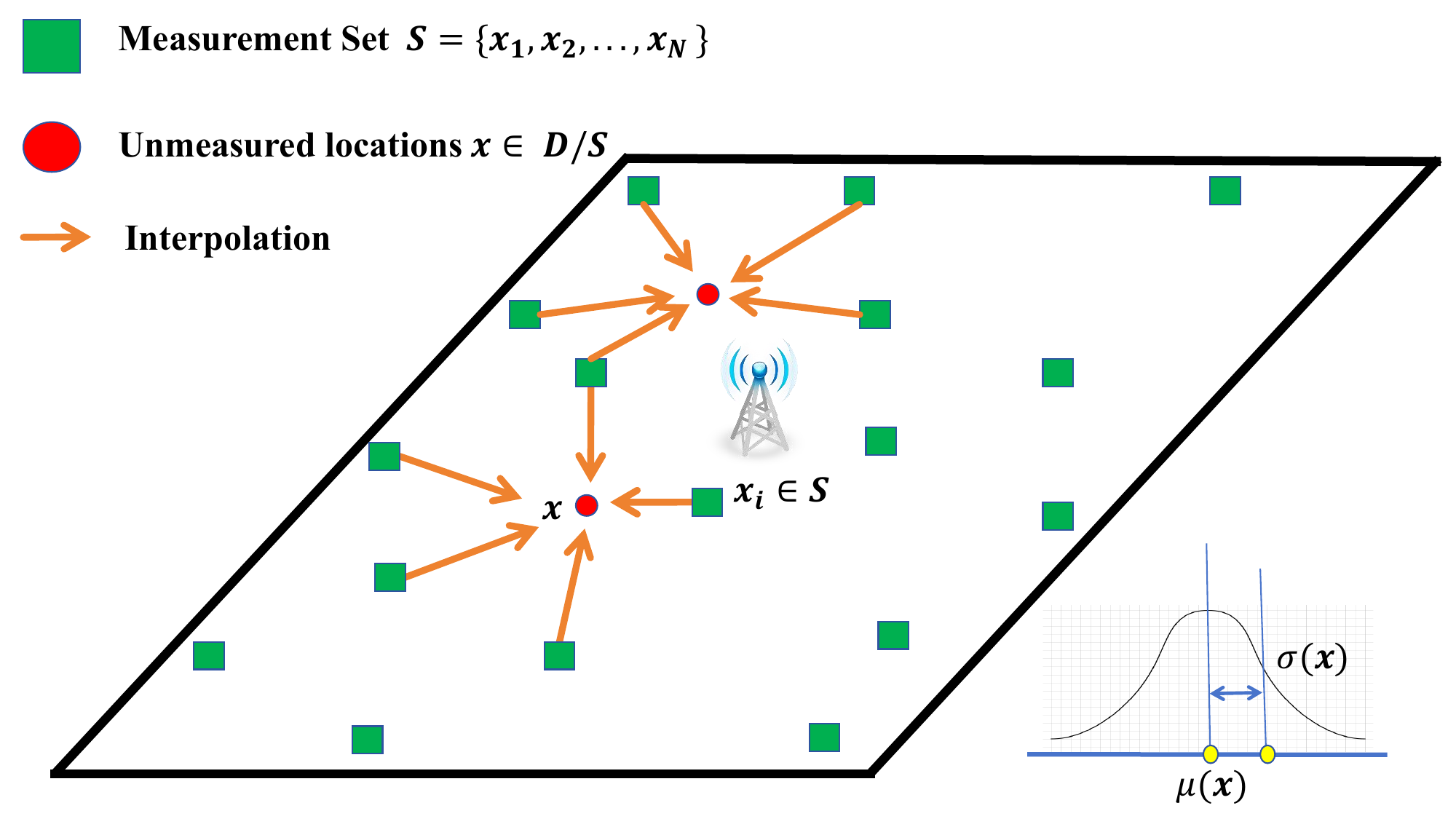}	
		\caption{An illustration of the measurement set $\boldsymbol{\mathcal{S}}$.}
		\label{fig:System Overview}
	\end{figure}

	As shown in Fig. \ref{fig:System Overview}, we aim to perform channel measurements at $N$ locations $\boldsymbol{\mathcal{S}}=\{\boldsymbol{x}_{1},\boldsymbol{x}_{2},...,\boldsymbol{x}_{N}\}\subseteq \boldsymbol{\mathcal{D}}$, and use the measurement data to infer the channel of those unmeasured locations $\boldsymbol{x} \in \boldsymbol{\mathcal{D}} \setminus \boldsymbol{\mathcal{S}}$, so that the AMSE is minimized. Specifically, we make two explicit assumptions as follows:

	\begin{enumerate}[label={\textbullet}]
		
		\item The mean $\mu(\boldsymbol{x})$ at each spatial location is known.
		
		\item The covariance function is known for all location-pairs.
		
	\end{enumerate}

	In this regime, we can separate the deterministic mean-value $\mu(\boldsymbol{x}_i)$ and the random fluctuating $S(\boldsymbol{x}_i)$ for each measured $\Gamma(\boldsymbol{x}_i),\boldsymbol{x}_i\in \boldsymbol{\mathcal{S}}$. Denote the estimation for each unmeasured ${\Gamma}(\boldsymbol{x}),\boldsymbol{x} \in \boldsymbol{\mathcal{D}}\setminus\boldsymbol{\mathcal{S}}$, as $\hat{\Gamma}(\boldsymbol{x})$, and the estimation for $S(\boldsymbol{x}),\boldsymbol{x} \in \boldsymbol{\mathcal{D}}\setminus\boldsymbol{\mathcal{S}}$, as $\hat{S}(\boldsymbol{x})$. Then, we have $\hat{\Gamma} (\boldsymbol{x})={\mu} (\boldsymbol{x})+\hat{S} (\boldsymbol{x})$, where $\hat{S}(\boldsymbol{x})$ is inferred based on those measurement values at the selected location set $\boldsymbol{\mathcal{S}}$, written as 
	
	\begin{equation}
		\hat{S} (\boldsymbol{x})= f\left( S(\boldsymbol{x}_{1}),  S(\boldsymbol{x}_{2}),..., S(\boldsymbol{x}_{N})  \right).
	\end{equation}

	 Define the MSE of each unmeasured $\Gamma(\boldsymbol{x}),\boldsymbol{x}\in \boldsymbol{\mathcal{D}}\setminus\boldsymbol{\mathcal{S}}$, as a function of the measurement set $\boldsymbol{\mathcal{S}}$, i.e.,

	\begin{align}
		\text{MSE}_{\boldsymbol{\mathcal{S}}} (\boldsymbol{x})= & \mathbb{E} \left[\left(\left( {\mu} (\boldsymbol{x})+\hat{S} (\boldsymbol{x})\right)-\left( \mu(\boldsymbol{x})+S(\boldsymbol{x})  \right)\right)^2\right] \notag \\
		= & \mathbb{E} \left[ \left( \hat{S} (\boldsymbol{x})-{S} (\boldsymbol{x}) \right)^2 \right],  \label{GeneralMSE}
	\end{align}where the expectation is taken with respect to the realizations of the random field. The AMSE of CGM in the continuous form is\footnote{Admittedly, such a formulation is not rigorous in the sense that $\text{MSE}_{\boldsymbol{\mathcal{S}}} (\boldsymbol{x})$ is non-integrable at the discontinuous zero-volume measured locations. That said, this simplified framing serves to highlight the generality of the underlying continuous framework.}

	\begin{equation}
		\text{AMSE}_{\boldsymbol{\mathcal{S}}} = \frac{1}{V_{\boldsymbol{\mathcal{D}}}}  \int_{\boldsymbol{x}\in \boldsymbol{\mathcal{D}}\setminus \boldsymbol{\mathcal{S}}} \text{MSE}_{\boldsymbol{\mathcal{S}}} (\boldsymbol{x}) \quad d \boldsymbol{x}, \label{Continuous_AMSE} 
	\end{equation} where $V_{\boldsymbol{\mathcal{D}}}$ denotes the volume of the considered 3D physical space. In order to make the AMSE in \eqref{Continuous_AMSE} more tractable, the continuous space is discretized into adjacent cubic grid points, where the channel gain at any location inside a specific grid point is characterized by the same random variable $\Gamma(\boldsymbol{x})$. At this point, $\boldsymbol{\mathcal{D}}$ is a finite set, and the global AMSE in this discrete framework is defined as 
	
	\begin{equation}
		\text{AMSE}_{\boldsymbol{\mathcal{S}}} = \frac{1}{|\boldsymbol{\mathcal{D}}|-|\boldsymbol{\mathcal{S}}|}  \sum_{\boldsymbol{x}\in \boldsymbol{\mathcal{D}}\setminus\boldsymbol{\mathcal{S}}} \text{MSE}_{\boldsymbol{\mathcal{S}}} (\boldsymbol{x}). \label{AMSE}
	\end{equation}

	We aim to determine the set of measurement locations $\boldsymbol{\mathcal{S}}$ so that the $\text{AMSE}_{\boldsymbol{\mathcal{S}}}$ of the remaining $\left(|\boldsymbol{\mathcal{D}}|-N\right)$ locations $\boldsymbol{\mathcal{D}}\setminus \boldsymbol{\mathcal{S}}$ is minimized, which can be formulated as
	
	\begin{subequations}
		\begin{align}
			\textup{(P1):}\quad 
			\min_{\boldsymbol{\mathcal{S}}}:\quad &  \text{AMSE}_{\boldsymbol{\mathcal{S}}} \\
			\text{s.t.} \quad &\boldsymbol{\mathcal{S}} \subseteq \boldsymbol {\mathcal{D}}, \\
			& |\boldsymbol{\mathcal{S}}| =N.
		\end{align} 
		\label{eq:1}
	\end{subequations}

	\section{Kriging-Based Variance Prediction}

	In order to determine the measurement set $\boldsymbol{\mathcal{S}}$, we need to have the closed-form expression of \eqref{GeneralMSE} for each unmeasured $\boldsymbol{x}$ to predict the global AMSE-reduction of measuring a specific set of locations. Since $S(\boldsymbol{x})$ is a zero-mean spatial field, the interpolation of unmeasured $\{S(\boldsymbol{x})|\boldsymbol{x} \in \boldsymbol{\mathcal{D}}\setminus\boldsymbol{\mathcal{S}} \}$ fits into the framework of ordinary Kriging, where a global constant mean-value is assumed. For each $\boldsymbol{x} \in \boldsymbol{\mathcal{D}}\setminus\boldsymbol{\mathcal{S}}$, define the set of $m$-nearest locations in $\boldsymbol{\mathcal{S}}$ as $\mathcal{N}_m(\boldsymbol{x}) \subseteq \boldsymbol{\mathcal{S}}$. In ordinary Kriging, each unmeasured $S(\boldsymbol{x})$ is estimated as the linear combination of the $m$ measured $\{S(\boldsymbol{x}_{n}) | \boldsymbol{x}_{n} \in \mathcal{N}_m(\boldsymbol{x}) \}$, i.e.,

	\begin{equation}
		\hat{S} (\boldsymbol{x}) = \sum_{\boldsymbol{x}_n \in \mathcal{N}_m(\boldsymbol{x})} \lambda_{n} {S}(\boldsymbol{x}_n). \label{Linear Combination}
	\end{equation}
	
	Denote the estimation error of ${S}(\boldsymbol{x})$ as $\epsilon_{\boldsymbol{x}}=\hat{S}(\boldsymbol{x})-{S}(\boldsymbol{x})$. Then, we have $\text{MSE}_{\boldsymbol{\mathcal{S}}} (\boldsymbol{x})=\mathbb{E}\left[\epsilon^2_{\boldsymbol{x}}\right]$. The variance of $\epsilon_{\boldsymbol{x}}$ is
	
	\begin{align}
		\sigma_{\epsilon_{\boldsymbol{x}}}^2=& \mathbb{E}\left[ \left( \epsilon_{\boldsymbol{x}}- \mathbb{E}\left[ \epsilon_{\boldsymbol{x}} \right]  \right)^2 \right] \notag \\
        =& \mathbb{E}\left[\epsilon_{\boldsymbol{x}}^2\right] -\mathbb{E}\left[\epsilon_{\boldsymbol{x}}\right]^2 \notag \\
		=& \text{MSE}_{\boldsymbol{\mathcal{S}}} (\boldsymbol{x})-\mathbb{E}\left[\epsilon_{\boldsymbol{x}}\right]^2.   \label{345345}
	\end{align}
	
	As can be seen from \eqref{345345}, when the estimator is unbiased, i.e.,
	
	\begin{equation}
		\mathbb{E}\left[\epsilon_{\boldsymbol{x}}\right]=\mathbb{E} \left[\hat{S}(\boldsymbol{x})-{S}(\boldsymbol{x})\right]= 0, \label{Unbiased}
	\end{equation} for all $\boldsymbol{x} \in \boldsymbol{\mathcal{D}}\setminus \boldsymbol{\mathcal{S}}$, we have

	\begin{equation}
		\sigma_{\epsilon_{\boldsymbol{x}}}^2= \text{MSE}_{\boldsymbol{\mathcal{S}}} (\boldsymbol{x})= \mathbb{E} \left[ \left(\hat{S}(\boldsymbol{x})-{S}(\boldsymbol{x}) \right)^2 \right]. \label{87644}
	\end{equation}

	Therefore, as long as the condition of unbiasedness \eqref{Unbiased} is satisfied, we can use the Kriging variance to approximate the MSE at each unmeasured location. By substituting \eqref{Linear Combination} into \eqref{Unbiased}, we have
	
	\begin{equation}
		\sum_{\boldsymbol{x}_n \in \mathcal{N}_m(\boldsymbol{x})} \lambda_{n} = 1 \label{sum=1}.   
	\end{equation}

	The covariance between $S(\boldsymbol{x}_i)$ and  $S(\boldsymbol{x}_j), \forall \boldsymbol{x}_i,\boldsymbol{x}_j \in \boldsymbol{\mathcal{S}}$ is $\mathbb{E} \left[ S(\boldsymbol{x}_i) S(\boldsymbol{x}_j) \right]=C_{ij}$ as defined in \eqref{covariance111}, and the covariance between the unmeasured $S(\boldsymbol{x})$ and $S(\boldsymbol{x}_n), \forall \boldsymbol{x}_n \in \mathcal{N}_m(\boldsymbol{x})$ is $\mathbb{E} \left[ S(\boldsymbol{x}_n) S(\boldsymbol{x}) \right]=C_{n0}$\footnote{Here, we use the notation ``0'' to represent the unmeasured location $\boldsymbol{x}$ that is being estimated, and the covariance $C_{n0}$ depends on the distance between the unmeasured $\boldsymbol{x}$ and $ \boldsymbol{x}_n \in \mathcal{N}_m(\boldsymbol{x})$.}. By substituting \eqref{Linear Combination} into \eqref{87644}, we have

	\begin{equation}
		\sigma_{\epsilon_{\boldsymbol{x}}}^2 =\sum_{\boldsymbol{x}_i \in \mathcal{N}_m(\boldsymbol{x})} \sum_{\boldsymbol{x}_j \in \mathcal{N}_m(\boldsymbol{x})} \lambda_{i} \lambda_{j} C_{ij}-2\sum_{\boldsymbol{x}_n \in  \mathcal{N}_m(\boldsymbol{x})} \lambda_{n} C_{n0} + \sigma^2(\boldsymbol{x}). \label{8654}
	\end{equation}
	
	Specifically, ordinary Kriging is based on the assumption of second-order stationarity, i.e., $\mathbb{E}[S(\boldsymbol{x})]=\mu=0$, and the covariance between two locations $\boldsymbol{x}_i$ and $\boldsymbol{x}_j$ depends only on their Euclidean distance $\|\boldsymbol{x}_i-\boldsymbol{x}_j\|$. In order to validate this assumption, we partition the universal set $\boldsymbol{\mathcal{D}}$ into $R$ disjoint subsets $\mathcal{D}_r,1\leq r\leq R$, corresponding to $R$ homogeneous subregions of the considered physical space. Correspondingly, we have $\boldsymbol{\mathcal{S}}=\mathcal{S}_1\cup \mathcal{S}_2\cup...\cup\mathcal{S}_R$. For $\boldsymbol{x}_i,\boldsymbol{x}_j\in \mathcal{D}_{r}$, we have $\sigma^2(\boldsymbol{x}_i)=\sigma^2(\boldsymbol{x}_j)=\sigma^2_r$, and $\mathcal{N}_m(\boldsymbol{x}) \subseteq \mathcal{S}_r, \boldsymbol{x} \in \mathcal{D}_{r}\setminus \mathcal{S}_r$. For each subregion, define the semivariogram as $\gamma_{ij}=\sigma^2_r-C_{ij}$\footnote{The covariance function, or equivalently, the semivariogram function is respectively fitted for each subregion. Here, we omit the subscript $r$ for simplicity.}. Then, \eqref{8654} can be expanded as

	\begin{align}
		\sigma_{\epsilon_{\boldsymbol{x}}}^2     =&\sum_{\boldsymbol{x}_i \in \mathcal{N}_m(\boldsymbol{x})} \sum_{\boldsymbol{x}_j \in \mathcal{N}_m(\boldsymbol{x})} \lambda_{i} \lambda_{j} \left(\sigma^2_r-\gamma_{ij}\right)\notag \\
        &-2\sum_{\boldsymbol{x}_n \in  \mathcal{N}_m(\boldsymbol{x})} \lambda_{n} \left(\sigma^2-\gamma_{n0} \right) + \sigma^2_r\notag \\
		=& \sigma^2_r \left(1+\sum_{\boldsymbol{x}_i \in \mathcal{N}_m(\boldsymbol{x})} \sum_{\boldsymbol{x}_j \in \mathcal{N}_m(\boldsymbol{x})} \lambda_{i} \lambda_{j} -2\sum_{\boldsymbol{x}_n \in  \mathcal{N}_m(\boldsymbol{x})} \lambda_{n} \right)
		\notag \\
		&+2 \sum_{\boldsymbol{x}_n \in \mathcal{N}_m(\boldsymbol{x})} \lambda_{n} \gamma_{n0} - \sum_{\boldsymbol{x}_i \in \mathcal{N}_m(\boldsymbol{x})} \sum_{\boldsymbol{x}_j \in \mathcal{N}_m(\boldsymbol{x})} \lambda_{i} \lambda_{j} \gamma_{ij}. 
	\end{align}  
	
	By using \eqref{sum=1}, we have
	
	\begin{equation}
		\sigma_{\epsilon_{\boldsymbol{x}}}^2  
		=2 \sum_{\boldsymbol{x}_n \in \mathcal{N}_m(\boldsymbol{x})} \lambda_{n} \gamma_{n0} - \sum_{\boldsymbol{x}_i \in \mathcal{N}_m(\boldsymbol{x})} \sum_{\boldsymbol{x}_j \in \mathcal{N}_m(\boldsymbol{x})} \lambda_{i} \lambda_{j} \gamma_{ij}. \label{67593}
	\end{equation}
	
	In order to tackle the equality constraint \eqref{sum=1}, a Lagrangian function is formulated as 
	
	\begin{equation}
		L\left(\lambda_{1},...,\lambda_{m}, \upsilon\right)=	\sigma_{\epsilon_{\boldsymbol{x}}}^2+ 2\upsilon \left( \sum_{\boldsymbol{x}_n \in \mathcal{N}_m(\boldsymbol{x})} \lambda_{n} - 1 \right),   \label{lagrangian} 
	\end{equation}where $\upsilon$ denotes the Lagrangian multiplier. By setting the derivatives of \eqref{lagrangian} with respect to each $\lambda_n$ to zero, we have $\frac{\partial L}{\partial \lambda_n} =2\gamma_{n0} -2 \sum_{j=1}^{m} \lambda_j \gamma_{nj}  + 2\upsilon = 0,  n = 1,2, \dots, m$, i.e.,
	
	\begin{equation}
		\sum_{\boldsymbol{x}_j \in \mathcal{N}_m(\boldsymbol{x})} \lambda_{j} \gamma_{nj} =\gamma_{n0}+\upsilon, \quad n = 1,2, \dots, m. \label{5464}
	\end{equation}
	
	By combining \eqref{sum=1} and \eqref{5464}, we have the matrix equation for ordinary Kriging as

	\begin{equation}
		\underbrace{
			\begin{bmatrix}
				\gamma_{11} & \gamma_{12} & \cdots & \gamma_{1m} & 1 \\
				\gamma_{21} & \gamma_{22} & \cdots & \gamma_{2m} & 1 \\
				\vdots & \vdots & \ddots & \vdots & \vdots \\
				\gamma_{m1} & \gamma_{m2} & \cdots & \gamma_{mm} & 1 \\
				1 & 1 & \cdots & 1 & 0
			\end{bmatrix}
		}_{\boldsymbol{R}^{\boldsymbol{\mathcal{S}}}}
		\underbrace{
			\begin{bmatrix}
				\lambda_{1} \\
				\lambda_{2} \\
				\vdots \\
				\lambda_{m} \\
				-\upsilon
			\end{bmatrix}
		}_{\boldsymbol{\Lambda}^{\boldsymbol{\mathcal{S}}}}
		=
		\underbrace{
			\begin{bmatrix}
				\gamma_{10} \\
				\gamma_{20} \\
				\vdots \\
				\gamma_{m0} \\
				1
			\end{bmatrix}
		}_{\boldsymbol{r}^{\boldsymbol{\mathcal{S}}}}. \label{matrix equation}
	\end{equation}

Note that the weighting coefficients $\boldsymbol{\Lambda}^{\boldsymbol{\mathcal{S}}}= {\left( \boldsymbol{R}^{\boldsymbol{\mathcal{S}}}\right)^{-1}} \boldsymbol{r}^{\boldsymbol{\mathcal{S}}}$. By substituting \eqref{5464} into \eqref{67593}, we have

	\begin{align}
		\sigma_{\epsilon_{\boldsymbol{x}}}^2 =&2 \sum_{\boldsymbol{x}_n \in \mathcal{N}_m(\boldsymbol{x})} \lambda_{n} \gamma_{n0}  - \sum_{\boldsymbol{x}_n \in \mathcal{N}_m(\boldsymbol{x})}\lambda_{n}  \left(  \sum_{\boldsymbol{x}_j \in \mathcal{N}_m(\boldsymbol{x})} \lambda_j \gamma_{nj} \right) \notag \\
		=&2 \sum_{\boldsymbol{x}_n \in \mathcal{N}_m(\boldsymbol{x})} \lambda_{n} \gamma_{n0} -\sum_{\boldsymbol{x}_n \in \mathcal{N}_m(\boldsymbol{x})}  \lambda_{n}  \left( \gamma_{n0}+\upsilon  \right) \notag \\
		=&2 \sum_{\boldsymbol{x}_n \in \mathcal{N}_m(\boldsymbol{x})} \lambda_{n} \gamma_{n0}-\sum_{\boldsymbol{x}_n \in \mathcal{N}_m(\boldsymbol{x})}\lambda_{n}  \gamma_{n0} -\sum_{\boldsymbol{x}_n \in \mathcal{N}_m(\boldsymbol{x})}  \lambda_{n} \upsilon    \notag \\
		=&\sum_{\boldsymbol{x}_n \in \mathcal{N}_m(\boldsymbol{x})}  \lambda_{n}\gamma_{n0}-\upsilon=\left(\boldsymbol{r}^{\boldsymbol{\mathcal{S}}} \right)^T\boldsymbol{\Lambda}^{\boldsymbol{\mathcal{S}}} \notag \\
        =&\left(\boldsymbol{r}^{\boldsymbol{\mathcal{S}}} \right)^T {\left( \boldsymbol{R}^{\boldsymbol{\mathcal{S}}}\right)^{-1}} \boldsymbol{r}^{\boldsymbol{\mathcal{S}}}. \label{ordinaryvariance}
	\end{align}

\section{Insight of Optimal Measurement Locations}

In this section, we analyze some special cases to gain some insight into the optimal measurement locations. For any two locations $\boldsymbol{x}_i,\boldsymbol{x}_j$, we consider an exponential semivariogram given as 
    
\begin{equation}
\gamma_{ij}(h_{ij}) =C_0+C(1-e^{-\frac{h_{ij}}{a}}), \label{Exponential} 
\end{equation}where $h_{ij}=\|\boldsymbol{x}_i -\boldsymbol{x}_j\|$ is the Euclidean distance between $\boldsymbol{x}_i$ and $\boldsymbol{x}_j$, $C_0$ is the nugget term, $a$ is the range of the semivariogram controlling the speed at which the spatial correlation decays, and $C_0+C$ is the sill value.

\subsection{Two Unmeasured Locations and One Measurement Location}

\begin{figure}[H]	
		\centering	
		\includegraphics[width=8cm,height=4.5cm]{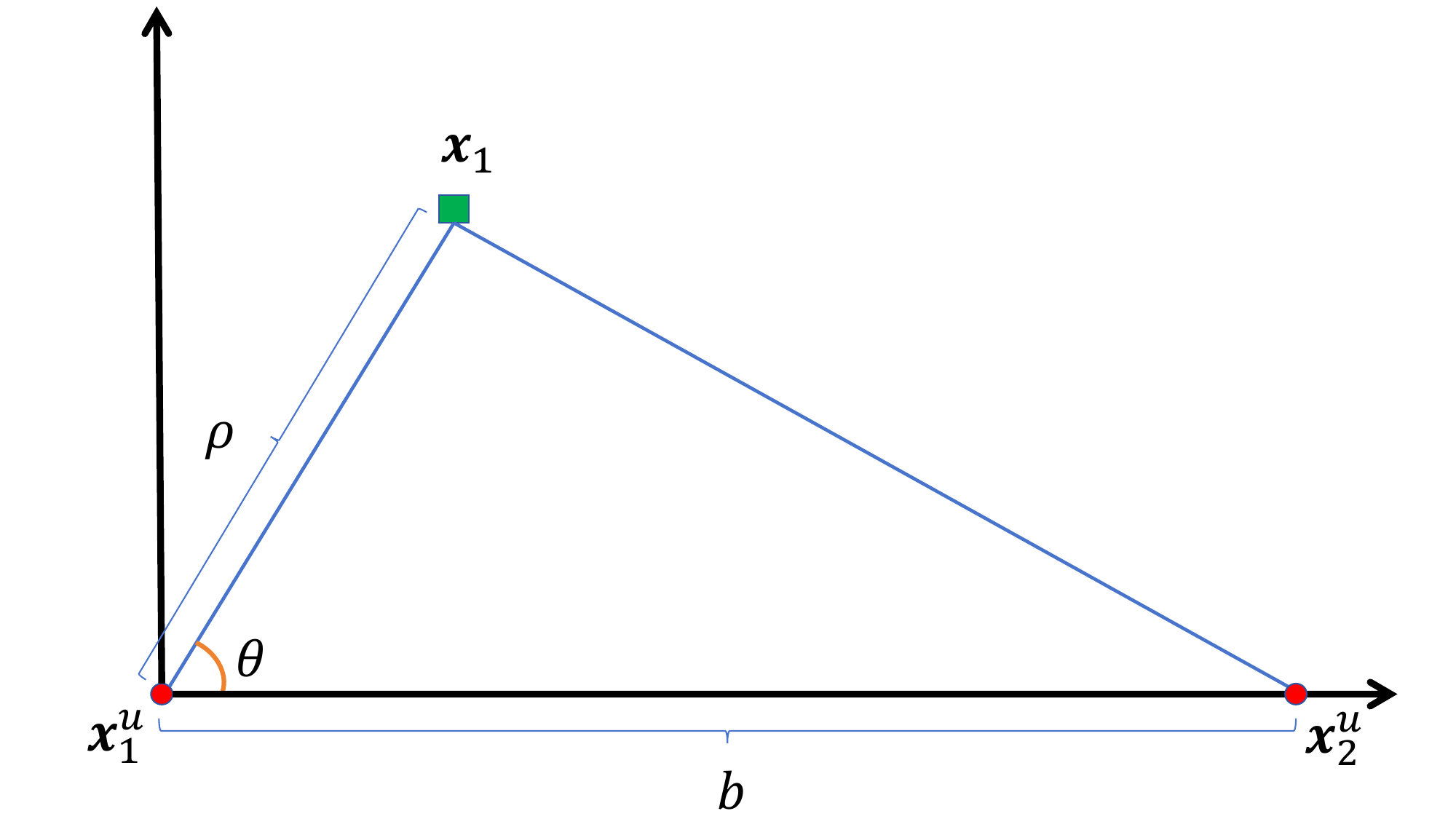}	
		\caption{A measurement scenario with two unmeasured locations and one measurement location.}
		\label{fig:2times1}
	\end{figure}

As illustrated in Fig. \ref{fig:2times1}, we consider the special case of one measurement location $\boldsymbol{x}_1$, and two unmeasured locations $\boldsymbol{x}^u_k,k=1,2$, where the superscript $u$ is employed to distinguish the unmeasured locations from the measurement location. Specifically, we use the polar coordinate system. The two unmeasured points are located at $(0,0)$ and $(b,0)$, and the coordinate of the measurement location is $\left(\rho,\theta\right)$.

Denote the semivariogram between the measurement location $\boldsymbol{x}_1$ and an unmeasured location $\boldsymbol{x}^u_k,k=1,2,$ as $\gamma_{1k}$. Using \eqref{ordinaryvariance}, we have the Kriging variance for $\boldsymbol{x}^u_k$ as 
	
	\begin{align}
		\sigma_{\epsilon_{\boldsymbol{x}}}^2=&\left(\boldsymbol{r}^{\boldsymbol{\mathcal{S}}} \right)^T {\left( \boldsymbol{R}^{\boldsymbol{\mathcal{S}}}\right)^{-1}} \boldsymbol{r}^{\boldsymbol{\mathcal{S}}}  \notag \\
		=&  \left[\gamma_{1k} \quad 1 \right] \begin{bmatrix}
			0 & 1 \\
			1  & 0
		\end{bmatrix}^{-1}
		\begin{bmatrix}
			\gamma_{1k} \\
			1
		\end{bmatrix}  \notag \\
		=&   \left[\gamma_{1k} \quad 1 \right] \begin{bmatrix}
			0 & 1 \\
			1  & 0
		\end{bmatrix}
		\begin{bmatrix}
			\gamma_{1k} \\
			1
		\end{bmatrix}  \notag \\
		=& 2\gamma_{1k}. \label{94654}
	\end{align}

Specifically, \eqref{94654} depends on the distance between $\boldsymbol{x}^u_k$ and the measurement point $\boldsymbol{x}_1$. Therefore, we can write the average Kriging variance of the two unmeasured locations as a function of the measurement point $\left(\rho,\theta\right)$. By substituting \eqref{Exponential} into \eqref{94654} and discarding the constant terms, we formulate a function to be minimized as

	\begin{equation}
		f\left(\rho,\theta\right)= -e^{-\frac{d_{11}}{a}}-  e^{-\frac{d_{12}}{a}},\rho\geq 0, 0\leq \theta<2\pi,
	\end{equation} where $d_{11}=\rho$, and $d_{12}=\sqrt{b^2+\rho^2-2b\rho \cos \theta}$. In order to find the global minimum of $f\left(\rho,\theta\right)$, we set the partial derivatives $\frac{\partial f}{\partial \theta}$ and $\frac{\partial f}{\partial \rho} $ to zero, and obtain

    	\begin{equation}  
		\left\{\begin{aligned}
     		&\frac{b\rho\sin \theta}{a\sqrt{b^2+\rho^2-2b\rho \cos \theta }} e^{-\frac{\sqrt{b^2+\rho^2-2b\rho \cos \theta }}{a}}=0,  \\
		& \frac{1}{a} \left(e^{-\frac{\rho}{a}} +\frac{\left(\rho-b\cos\theta\right)}{\sqrt{b^2+\rho^2-2b\rho \cos \theta }}e^{-\frac{\sqrt{b^2+\rho^2-2b\rho \cos \theta }}{a}} \right)=0.  
        		\end{aligned}\right. 
	\end{equation}

\begin{figure}[H]	
		\centering	
		\includegraphics[width=7cm,height=4.5cm]{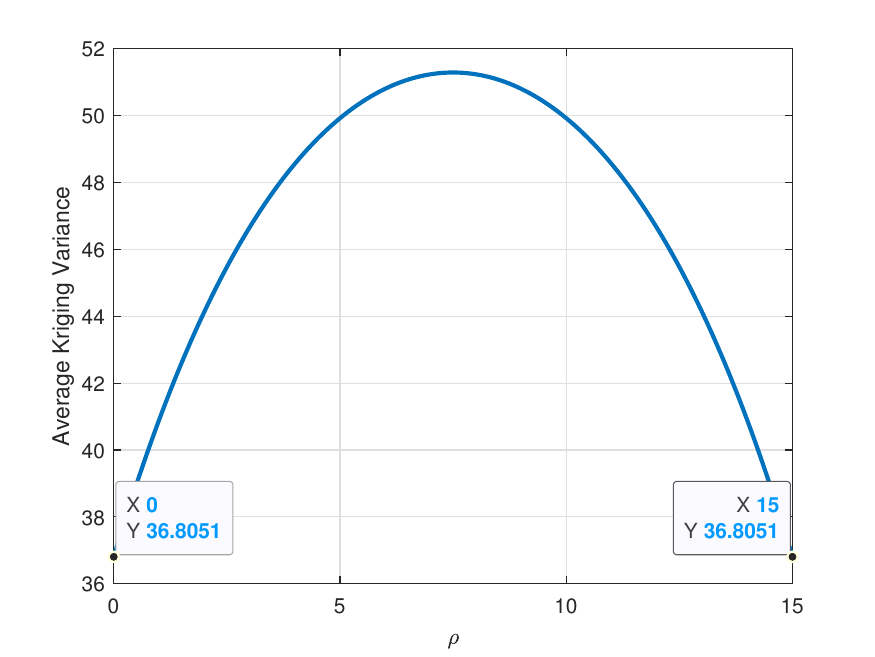}	
		\caption{The average Kriging variance of two unmeasured locations, where $a=5\, \text{m}$, $b=15\,\text{m}$, $C_0=12$, $C=48$, and $\theta =0$.}
		\label{fig:2times1_function}
	\end{figure}

The only solution to the above equations is $\rho=\frac{b}{2}, \theta=0$. Therefore, $\left( \frac{b}{2} ,0\right)$ is the only stationary point of $f\left(\rho,\theta\right)$. The extrema discriminant at $\left( \frac{b}{2} ,0\right)$ is $\frac{\partial^2 f}{\partial \rho^2}\frac{\partial^2 f}{\partial \theta^2}-\frac{\partial^2 f}{\partial \rho\partial \theta}=- \frac{2b}{a^3} e^{-\frac{b}{2a}} <0$, and therefore $\left( \frac{b}{2},0\right)$ is a saddle point instead of a global extremum. Intuitively, the global minimum should be located on the line between the two unmeasured locations, i.e., $\theta=0$, as it minimizes $d_{11}+d_{12}$. As shown in Fig. \ref{fig:2times1_function}, when $\theta =0$, the minimum average Kriging variance is achieved at $(0,0)$ and $(b,0)$, corresponding to the locations of the two unmeasured points, while the local maximum is achieved at $\left(\frac{b}{2},0\right)$. The implication of the result is that when there are two unmeasured locations and one measurement location, the optimal measurement strategy is to measure one of the two unmeasured points while sacrificing the estimation accuracy of the other.

\subsection{Three Unmeasured Locations and One Measurement Location}

	\begin{figure}[H]	
		\centering	
		\includegraphics[width=8cm,height=4.5cm]{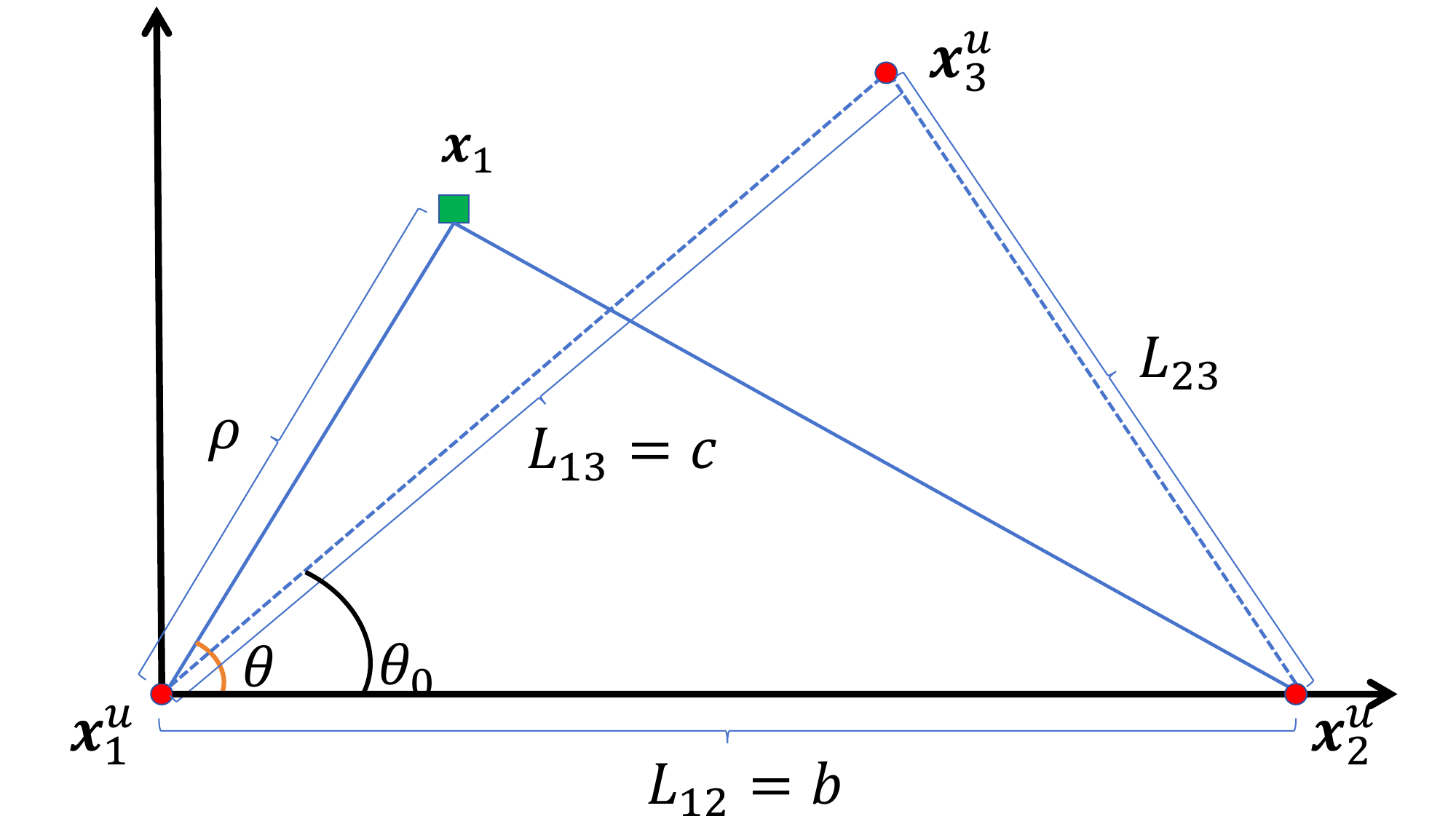}	
		\caption{A measurement scenario with three unmeasured locations and one measurement location.}
		\label{fig:3times1}
	\end{figure}
    
Consider the special case of three unmeasured locations and one measurement location as illustrated in Fig. \ref{fig:3times1}. The coordinates of the three unmeasured locations are $(0,0)$, $(b,0)$, $(c,\theta_0)$, and the coordinate of the measurement location is $\left(\rho,\theta\right)$.	Similarly, we formulate a function to be minimized as
	
	\begin{align}
		f\left(\rho,\theta\right)=  -e^{-\frac{d_{11}}{a}}-  e^{-\frac{d_{12}}{a}} - e^{-\frac{d_{13}}{a}},\rho\geq 0, 0\leq \theta<2\pi, \label{96758373111}
	\end{align} where $d_{11}=\rho$ $d_{12}=\sqrt{b^2+\rho^2-2b\rho \cos \theta}$ $d_{13}=\sqrt{c^2+\rho^2-2c\rho \cos \left( \theta- \theta_0  \right)}$. In order to find the global minimum of $f\left(\rho,\theta\right)$, we set the partial derivatives $\frac{\partial f}{\partial \theta}$ and $\frac{\partial f}{\partial \rho} $ to zero, and obtain

	\begin{equation}  
		\left\{\begin{aligned}
			&\frac{c\sin \left(\theta-\theta_0\right)}{\sqrt{c^2+\rho^2-2c\rho \cos \left(\theta-\theta_0\right)}}e^{-\frac{\sqrt{c^2+\rho^2-2c\rho \cos \left(\theta-\theta_0\right) }}{a}}  \\
			&  +\frac{\rho\sin\theta}{\sqrt{b^2+\rho^2-2b\rho \cos \theta }}e^{-\frac{\sqrt{b^2+\rho^2-2b\rho \cos \theta }}{a}} =0, \\
			&e^{-\frac{\rho}{a}}+ \frac{\rho-b\cos\theta}{\sqrt{b^2+\rho^2-2b\rho \cos \theta }}e^{-\frac{\sqrt{b^2+\rho^2-2b\rho \cos \theta }}{a}} \\
			&+\frac{\rho-c\cos \left(\theta-\theta_0\right)}{\sqrt{c^2+\rho^2-2c\rho \cos \left(\theta-\theta_0\right)}}e^{-\frac{\sqrt{c^2+\rho^2-2c\rho \cos \left(\theta-\theta_0\right) }}{a}} 
			=0.
		\end{aligned}\right. 
		\label{super}
	\end{equation}

\begin{figure}[H]	
		\centering	
		\includegraphics[width=8cm,height=5cm]{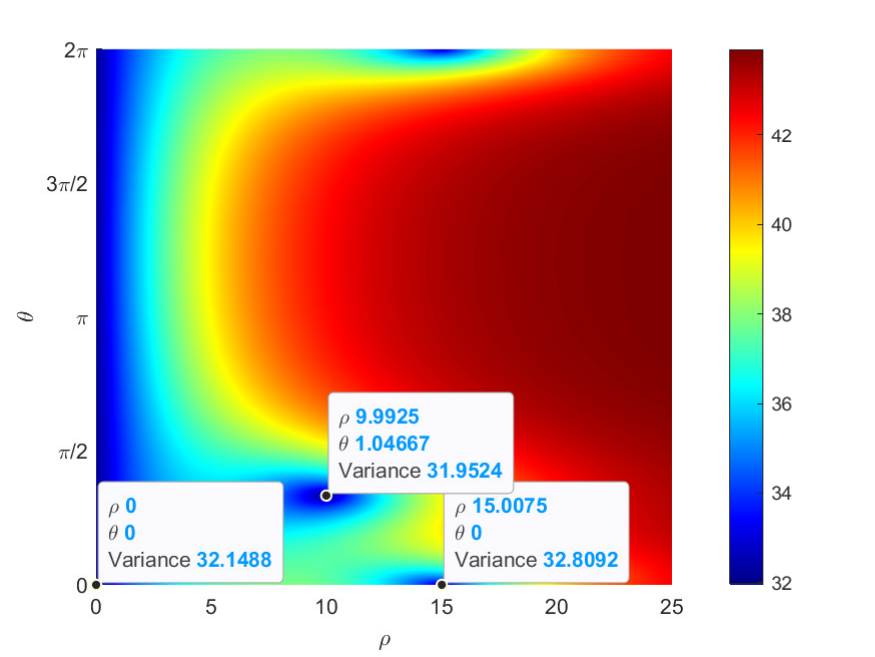}	
		\caption{The average Kriging variance of three unmeasured locations, where $a=5\,\text{m}$, $b=15\,\text{m}$, $c=10\,\text{m}$, $\theta_0=\frac{\pi}{3}$, $C_0=12$, and $C=48$.}
		\label{fig:3times1_function}
	\end{figure}

Note that \eqref{super} is a set of transcendental equations without closed-form solutions, and we can use the Newton-Raphson method to find the stationary points of $f\left(\rho,\theta\right)$. For the geometry given in Fig. \ref{fig:3times1}, the pairwise distances among the three unmeasured locations are $L_{12}=b$, $L_{13}=c$, and $L_{23}=\sqrt{b^2+c^2-2bc\cos \theta_0}$. For any location $\left(\rho,\theta\right)$ that is not at the three unmeasured locations, when the range value $a< \min \{L_{12},L_{13},L_{23} \}$, $f\left(\rho,\theta\right)$ is dominated by $-e^{-\frac{d_{1k_{\text{nearest}}}}{a}}$, where $k_{\text{nearest}}=\arg \min_k d_{1k}$. Meanwhile, due to the monotonicity of the exponential function, $-e^{-\frac{d_{1k_{\text{nearest}}}}{a}}$ decreases monotonically as $d_{1k_{\text{nearest}}}$ decreases. Therefore, the local minima of $f\left(\rho,\theta\right)$ are at the three unmeasured locations, as shown in Fig. \ref{fig:3times1_function}. In this regime, the optimal measurement location is the unmeasured location $\boldsymbol{x}^u_{k_{\text{best}}}$, where

\begin{equation}
 k_{\text{best}}=\arg \min_k  \sum_{y\in \{1,2,3\}\setminus \{k\}} -e^{-\frac{d_{1y}}{a}}.   
\end{equation}

\begin{figure}[H]	
		\centering	
		\includegraphics[width=8cm,height=5cm]{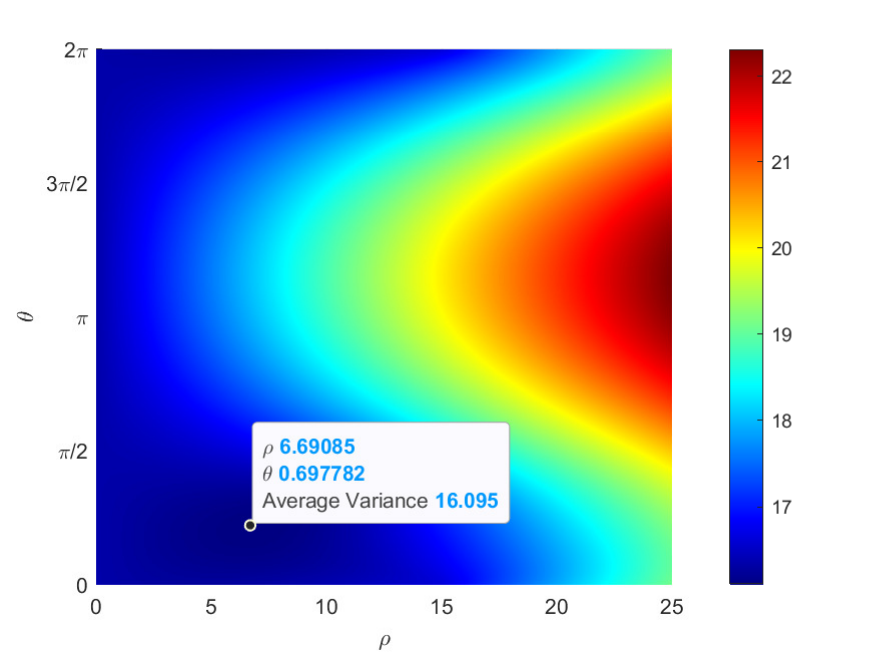}	
		\caption{The average Kriging variance of three unmeasured locations, where $a=100\,\text{m}$, $b=15\,\text{m}$, $c=10\,\text{m}$, $\theta_0=\frac{\pi}{3}$, $C_0=12$, and $C=48$.}
		\label{fig:3times1_}
	      \end{figure}

On the other hand, when $a\gg \max \{L_{12},L_{13},L_{23} \}$, the distance-dependent attenuation of each exponential component becomes sufficiently small and comparable to each other. In this regime, the three unmeasured locations reduce to a group of saddle points, and the genuine global minima is the stationary point as the solution to \eqref{super}. As shown in Fig. \ref{fig:3times1_}, the global minimum corresponds to the stationary point in the interior of the triangle formed by the unmeasured points, where the exponential contributions of the three unmeasured locations are synergistically balanced.

	\subsection{Two Measurement Locations}

Consider the special case of two measurement locations $\boldsymbol{x}_1,\boldsymbol{x}_2$, and three unmeasured location $\boldsymbol{x}^u_k,k=1,2,3$. Denote the semivriogram between an unmeasured location $\boldsymbol{x}^u_k$ and the measurement locations $\boldsymbol{x}_1,\boldsymbol{x}_2$, as $\gamma_{1k}$ and $\gamma_{2k}$. Using \eqref{ordinaryvariance}, we have the Kriging variance for $\boldsymbol{x}^u_k$ as\footnote{Note that the $\gamma_{12}$ in \eqref{84756343} refers to the semivariogram between the two measurement locations.}  
	
	\begin{align}
		\sigma_{\epsilon_{k}}^2=&\left(\boldsymbol{r}^{\boldsymbol{\mathcal{S}}} \right)^T {\left( \boldsymbol{R}^{\boldsymbol{\mathcal{S}}}\right)^{-1}} \boldsymbol{r}^{\boldsymbol{\mathcal{S}}}  \notag \\
		=&  \begin{bmatrix}
			\gamma_{1k}& \gamma_{2k} & 1 
		\end{bmatrix}
		\begin{bmatrix}
			0 & \gamma_{12} & 1 \\
			\gamma_{12} & 0 &   1 \\
			1 & 1 & 0
		\end{bmatrix}^{-1}
		\begin{bmatrix}
			\gamma_{1k} \\
			\gamma_{2k} \\
			1
		\end{bmatrix}  \notag \\
		=& \frac{1}{2\gamma_{12}} \begin{bmatrix}
			\gamma_{1k}& \gamma_{2k} & 1 
		\end{bmatrix}\begin{bmatrix}
			-1 & 1 & \gamma_{12} \\
			1 & -1 &  \gamma_{12} \\
			\gamma_{12} & \gamma_{12} & -\gamma_{12}^2
		\end{bmatrix}
		\begin{bmatrix}
			\gamma_{1k} \\
			\gamma_{2k} \\
			1
		\end{bmatrix}  \notag \\
		=& \frac{2\gamma_{12}\gamma_{1k}+2\gamma_{12}\gamma_{2k}+2\gamma_{1k}\gamma_{2k}-\gamma_{1k}^2-\gamma_{2k}^2-\gamma_{12}^2}{2\gamma_{12}} \notag \\
		=& \gamma_{1k}+\gamma_{2k}-\frac{\gamma_{12}}{2}-\frac{\left(\gamma_{1k}-\gamma_{2k} \right)^2}{2\gamma_{12}}.\label{84756343}
	\end{align} 
	
By substituting \eqref{Exponential} into \eqref{84756343} and discarding the constant terms, we formulate a function to be minimized as

	\begin{align}
		&f\left(\rho_1,\rho_2,\theta_1,\theta_2\right)= \notag \\
          -&e^{-\frac{d_{11}}{a}}-  e^{-\frac{d_{21}}{a}} - e^{-\frac{d_{12}}{a}}-e^{-\frac{d_{22}}{a}}- e^{-\frac{d_{13}}{a}}-e^{-\frac{d_{23}}{a}}  \notag \\
		-& \frac{C\left( \left(e^{-\frac{d_{11}}{a}}-  e^{-\frac{{d_{21}}}{a}} \right)^2+\left(e^{-\frac{d_{12}}{a}}-  e^{-\frac{d_{22}}{a}} \right)^2+\left(e^{-\frac{d_{13}}{a}}-  e^{-\frac{d_{23}}{a}} \right)^2 \right)}{2\left( C_0+C\left(1-e^{-\frac{d_{12}^M}{a}} \right) \right)} \notag \\
        +&\frac{3}{2} e^{-\frac{d_{12}^M}{a}}, h_1,h_2\geq 0, 0\leq\theta_1,\theta_2 < 2\pi, \label{Four}
	\end{align}	where $d_{i1}=\rho_i, i=1,2$, $d_{i2}=\sqrt{b^2+\rho_i^2-2b\rho_i \cos \theta_i}, i=1,2$, $d_{i3}=\sqrt{c^2+\rho_i^2-2c\rho_i \cos  \left( \theta_i-\theta_0\right)}, i=1,2$, and $d_{12}^M=\sqrt{\rho_1^2+\rho_2^2-2\rho_1\rho_2\cos \left( \theta_1-\theta_2\right)}$.

	\begin{figure}[H]	
		\centering	
		\includegraphics[width=7cm,height=4.5cm]{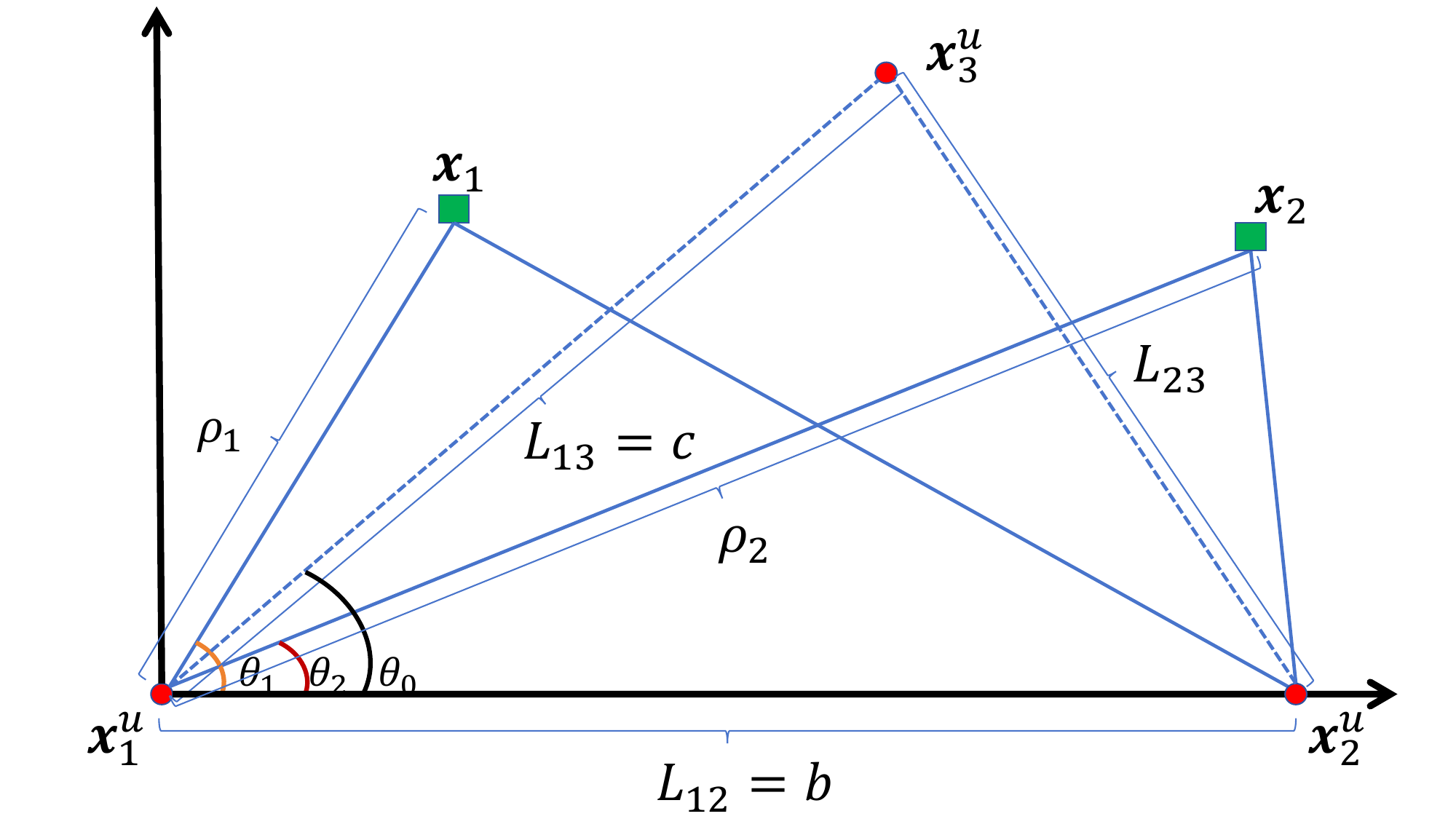}	
		\caption{A measurement scenario with three unmeasured locations and two measurement locations.}
		\label{fig:3times2}
	\end{figure}

Similarly to the single measurement location scenario, when the range value $a< \min \{L_{12},L_{13},L_{23} \}$, the exponential components $e^{-\frac{d_{ik}}{a}}, i=1,2;k=1,2,3$ dominate the value of \eqref{Four}, and therefore the two measurement locations tend to be attracted to two of the three unmeasured locations. In the following, we determine which pair of the unmeasured locations would be selected. With a slight abuse of notation, we denote the three unmeasured locations $\boldsymbol{x}^u_k,k=1,2,3$ by $o,p,q$. By substituting the above notation into \eqref{Four}, we have the function value when measuring the locations $o$ and $p$ as

\begin{align}
    f_{op}=&\underbrace{- \left( 2+\frac{1}{2} e^{-\frac{L_{op}}{a}}+e^{-\frac{L_{oq}}{a}}+e^{-\frac{L_{pq}}{a}} \right)}_{z_{op}} \notag \\
    & \underbrace{-\frac{C\left( \left(1-e^{-\frac{L_{op}}{a}} \right)^2 + \left( e^{-\frac{L_{oq}}{a}}-e^{-\frac{L_{pq}}{a}}  \right)^2  \right)}{2\left( C_0+C\left(1-e^{-\frac{L_{op}}{a}} \right) \right)}.}_{t_{op}}
\end{align}

First, we compare the exponential terms $z_{12},z_{13},z_{23}$. Suppose that $L_{12}>L_{13}>L_{23}$. Then, we have $z_{12}-z_{13}=\frac{1}{2} \left(e^{-\frac{L_{12}}{a}}-e^{-\frac{L_{13}}{a}} \right)<0$, and $z_{12}-z_{23}=\frac{1}{2} \left(e^{-\frac{L_{12}}{a}}-e^{-\frac{L_{23}}{a}} \right)<0$. In terms of the fractional term $t_{op}$, when $L_{op}\gg a$, the term $\left(1-e^{-\frac{L_{op}}{a}} \right)^2$ dominates the numerator of $t_{op}$, and therefore $t_{op}$ can be reduced to $t_{op} \approx-\frac{C \left(1-e^{-\frac{L_{op}}{a}} \right)^2   }{2\left( C_0+C\left(1-e^{-\frac{L_{op}}{a}} \right) \right)}$. Let $1-e^{-\frac{L_{op}}{a}}=u$. We formulate a function $t(u)= -\frac{Cu^2}{C_0+Cu}$. For $u=1-e^{-\frac{L_{op}}{a}}>0$, we have $t'(u)=-\frac{2CC_0u+C^2u^2}{\left(C_o+Cu \right)^2}<0$. A larger $L_{op}$ yields a larger $u$, hence a smaller $t_{op}$. 

In summary, when the range value $a< \min \{L_{12},L_{13},L_{23} \}$, the spatial correlation between any pair of unmeasured locations becomes negligible. Consequently, the optimal measurement strategy is to directly measure the pair of unmeasured locations with the maximum spatial distance. This strategy maximizes the spatial coverage of the measurement set $\boldsymbol{\mathcal{S}}$, and delivers the greatest reduction in the average Kriging variance. On the other hand, when $a\gg \max \{L_{12},L_{13},L_{23} \}$, the three unmeasured locations become highly correlated, and the optimal measurement locations are located within the triangle formed by the unmeasured locations.

\section{Proposed Algorithms}
	
	In this section, we present two algorithms to solve problem (P1) given in \eqref{eq:1} in the general case with arbitrary number of measurement locations.

	\subsection{Greedy Search}
	
	The greedy algorithm adopts the strategy of greedily selecting the measurement location that yields the highest AMSE-reduction among all candidate locations at each iteration. This algorithm is intuitively straightforward and easy to implement. The pseudo-code is given as follows.
	
	\begin{breakablealgorithm}
		\caption{Greedy algorithm for solving problem (P1)}
		\begin{algorithmic}[1] 
			\State \textbf{Input:} The global universal set $\boldsymbol{\mathcal{D}}$. $\boldsymbol{\mathcal{S}} \gets \emptyset$.
			\For{$n \gets 1$ to $N$}  
			\For{each $\boldsymbol{s}_i\in \boldsymbol{\mathcal{D}}\setminus \boldsymbol{\mathcal{S}}$}  
			\State $\text{AMSE}_{n,i}=\frac{1}{|\boldsymbol{\mathcal{D}}|-|\boldsymbol{\mathcal{S}}|-1}  \sum_{\boldsymbol{x}\in \boldsymbol{\mathcal{D}}\setminus \boldsymbol{\mathcal{S}}\setminus \boldsymbol{s}_i} \text{MSE}_{\boldsymbol{\mathcal{S}}\cup \{ \boldsymbol{s}_i\}}(\boldsymbol{x})$.
			\EndFor 
			\State $m \leftarrow \arg\min_{i \in \{1,2,...|\boldsymbol{\mathcal{D}}\setminus \boldsymbol{\mathcal{S}}|\}} \text{AMSE}_{n,i}$.
			\State $\boldsymbol{\mathcal{S}}\leftarrow \boldsymbol{\mathcal{S}} \cup\{\boldsymbol{s}_m\}$.
			\State $n\leftarrow n+1$.
			\EndFor 
			\State \Return $\boldsymbol{\mathcal{S}}$.
		\end{algorithmic}
	\end{breakablealgorithm}

	For the greedy algorithm, the number of iterations of the outer loop is the size of the measurement set $\boldsymbol{\mathcal{S}}$, i.e., $N$. In the $n$-th outer loop, we have $|\boldsymbol{\mathcal{S}}|=n-1,1\leq n \leq N$, and the number of candidate measurement locations is $|\boldsymbol{\mathcal{D}}\setminus \boldsymbol{\mathcal{S}}|=|\boldsymbol{\mathcal{D}}|-(n-1), 1\leq n \leq N$. In order to determine the optimal measurement location in the $n$-th outer loop, we need to evaluate the updated AMSE, i.e., $\text{AMSE}_{n,i}$, of measuring the unmeasured location $\boldsymbol{s}_i\in \boldsymbol{\mathcal{D}}\setminus \boldsymbol{\mathcal{S}}$. Therefore, the total number of inner iterations in the $n$-th outer loop is $|\boldsymbol{\mathcal{D}}|-(n-1), 1\leq n \leq N$. In order to determine $\text{AMSE}_{n,i}$, we need to calculate the Kriging variance of the remaining $|\boldsymbol{\mathcal{D}}|-(n-2),1\leq n \leq N$, unmeasured locations. For the Kriging variance of each unmeasured location, the main computation is on the inversion of the coefficient matrix $\boldsymbol{R}^{\boldsymbol{\mathcal{S}}}$ in \eqref{ordinaryvariance}, which is $(m+1)^3$. Given that $|\boldsymbol{\mathcal{D}}|\gg N$, we approximate the number of inner iterations in the $n$-th outer loop, i.e., $|\boldsymbol{\mathcal{D}}|-(n-1)$, and the number of unmeasured locations for the calculation of $\text{AMSE}_{n,i}$, i.e., $|\boldsymbol{\mathcal{D}}|-(n-2)$, as $|\boldsymbol{\mathcal{D}}|$. Therefore, the overall time-complexity of the greedy algorithm is $\mathcal{O}\left(N|\boldsymbol{\mathcal{D}}|^2(m+1)^3\right)$.

	\subsection{Simulated Annealing}
	
	Despite its simplicity, the greedy algorithm is prone to local optima. In order to explore the lower bound of \eqref{eq:1}, we employ the Monte Carlo SA. As a heuristic algorithm, SA draws on the principle of thermodynamic annealing \cite{SA} in physics. An important component of SA is the probabilistic acceptance rule known as the Metropolis criterion. At initial high temperatures, the Metropolis criterion allows SA to accept suboptimal solutions in a probabilistic manner, promoting the exploration of unexplored spatial subregions, which effectively avoids local optima. With progressive temperature annealing, the algorithm evolves from global probabilistic Metropolis exploration to local deterministic greedy exploitation, converging to the globally optimal measurement pattern. The pseudo-code is summarized as follows.
	
	\begin{breakablealgorithm}
		\caption{Simulated annealing for solving problem (P1)}
		\begin{algorithmic}[1] 
			\State \textbf{Initialize:} Initial temperature $T_0$, termination temperature $T_t$, cooling rate $\alpha$, number of element swap at each iteration $N_{\text{swap}}$, $\boldsymbol{\mathcal{S}}_{\text{best}} \gets \emptyset$.
			
			\State Randomly generate an initial set $\boldsymbol{\mathcal{S}}_0 = \{\boldsymbol{x}_1, \boldsymbol{x}_2, ..., \boldsymbol{x}_N\}$, where $\boldsymbol{x}_i \in \boldsymbol{\mathcal{D}},1\leq i\leq N$, $\boldsymbol{\mathcal{S}}_{\text{best}} \gets \boldsymbol{\mathcal{S}}_0$.
			\State $T  \gets T_0$, $K \leftarrow \left\lfloor \frac{\ln(T_t/T_0)}{\ln \alpha} \right\rfloor+1$
			
			\For {$k = 1$ to $K$}
			\State $T_k \leftarrow T_0 \cdot \alpha^{k-1}$
			\For {i = 1 to $N_{\text{swap}}$ }
			\State Calculate $\mathcal{J}_{\text{current}} =\frac{1}{|\boldsymbol{\mathcal{D}}|-N}  \sum_{\boldsymbol{x}\in \boldsymbol{\mathcal{D}}\setminus\boldsymbol{\mathcal{S}}_{\text{best}}} \text{MSE}_{\boldsymbol{\mathcal{S}}_{\text{best}}}(\boldsymbol{x})$. 
			
			\State Randomly select $\boldsymbol{x}\in \boldsymbol{\mathcal{S}}_{\text{best}}$, $\boldsymbol{y}\in \boldsymbol{\mathcal{D}}\setminus \boldsymbol{\mathcal{S}}_{\text{best}}$.
			
			\State $\boldsymbol{\mathcal{S}}_{\text{best}} =\boldsymbol{\mathcal{S}}_{\text{best}}\setminus \{\boldsymbol{x} \}\cup \{\boldsymbol{y}\}$.
			
			\State Calculate $\mathcal{J}_{\text{new}} = \frac{1}{|\boldsymbol{\mathcal{D}}|-N}  \sum_{\boldsymbol{x}\in \boldsymbol{\mathcal{D}}\setminus\boldsymbol{\mathcal{S}}_{\text{best}}} \text{MSE}_{\boldsymbol{\mathcal{S}}_{\text{best}}}(\boldsymbol{x})$. 
			
			\If{$\mathcal{J}_{\text{current}} < \mathcal{J}_{\text{new}}$}
			\State $\Delta \mathcal{J} = \mathcal{J}_{\text{new}} - \mathcal{J}_{\text{current}}$. 
			\State Acceptance probability $P = \exp\left(-\frac{\Delta \mathcal{J}}{T}\right)$.  
			\State Generate a random number $z \sim \text{Uniform}(0, 1)$.
			\If{$z > P$}
			\State $\boldsymbol{\mathcal{S}}_{\text{best}} =\boldsymbol{\mathcal{S}}_{\text{best}}\setminus \{\boldsymbol{y} \}\cup \{\boldsymbol{x}\}$.
			\EndIf
			\EndIf
			\EndFor 
			\EndFor			
			\State \Return $\boldsymbol{\mathcal{S}}_{\text{best}}$.
		\end{algorithmic}
	\end{breakablealgorithm}

	For the SA, the number of iterations of the outer loop is $K= \left\lfloor \frac{\ln(T_t/T_0)}{\ln \alpha} \right\rfloor+1$, and each outer loop has $N_{\text{swap}}$ rounds of inner iterations. For the calculation of the updated AMSE, each inner loop needs to perform $|\boldsymbol{\mathcal{D}}|$ times of coefficient matrix inversion. Therefore, the overall time-complexity of the proposed SA is $\mathcal{O}\left(K N_{\text{swap}}|\boldsymbol{\mathcal{D}}| (m+1)^3\right)$.

\section{Adaptive Discretization}


In the formulation of (P1), we assumed that the underlying physical space is uniformly discretized into the finite set $\boldsymbol{\mathcal{D}}$ with a global uniform discretization granularity $\Delta_0$. In order for CGM to fully capture the channel gain variation in complex physical environments, the discretization granularity $\Delta_0$ needs to be small enough so that the distance between the centers of adjacent grid points does not exceed the local spatial correlation distance. Nevertheless, such a restriction poses extra computational burden for determining the measurement set $\boldsymbol{\mathcal{S}}$. As pointed out in Section V, the time-complexities of both the greedy algorithm and the SA depend on $|\boldsymbol{\mathcal{D}}|$. Specifically, $|\boldsymbol{\mathcal{D}}|$ is proportional to $\Delta_0^{-3}$, and a linear decrease in $\Delta_0$ leads to a cubic increase in algorithmic complexity, rendering \eqref{eq:1} prohibitively complex when $\Delta_0$ is small.

In order to make \eqref{eq:1} more computationally tractable, we delve into the computational complexities of the two proposed algorithms in Section V. For the greedy algorithm, the number of outer iterations, i.e., $N$, and the matrix inversion complexity $\left( m+1 \right)^3$, are unchangeable. Meanwhile, in order to ensure the optimality of the selected measurement locations, the calculation of each predicted $\text{AMSE}_{n,i}$ must be performed with respect to the global universal set $\boldsymbol{\mathcal{D}}$. Therefore, the only way to reduce the computational complexity of the greedy algorithm is to reduce the number of candidate measurement locations from $\boldsymbol{\mathcal{D}}$, to a smaller set $\boldsymbol{\mathcal{U}}\subseteq \boldsymbol{\mathcal{D}}$, where $|\boldsymbol{\mathcal{U}}|=U\ll |\boldsymbol{\mathcal{D}}|$. For the SA, in order to ensure the effectiveness of element swaps in each outer round, the number of swaps $N_{\text{swap}}$ must be of the same order as the number of candidate measurement locations. Similarly, we can reduce the candidate location set to a smaller set $\boldsymbol{\mathcal{U}}\subseteq \boldsymbol{\mathcal{D}}$ to correspondingly reduce the number of element swaps in each outer iteration. To sum up, reducing the candidate set to a subset of all possible unmeasured locations effectively decreases the computational complexities of both algorithms, which raises the question of how to design the reduced candidate set $\boldsymbol{\mathcal{U}}$.

As introduced in Section III, we partition the physical space into $R$ homogeneous subregions by dividing the universal set $\boldsymbol{\mathcal{D}}$ into $R$ disjoint subsets $\mathcal{D}_r,1\leq r\leq R$, to validate the second-order stationarity assumption. In this section, we develop an adaptive discretization strategy to adaptively allocate the $|\boldsymbol{\mathcal{U}}|=U$ grid points across the $R$ subregions by adjusting the discretization granularity of each subregion. Correspondingly, the reduced candidate set $\boldsymbol{\mathcal{U}}$ is partitioned into $R$ disjoint subsets $\mathcal{U}_r \subseteq \mathcal{D}_r,1\leq r\leq R$. In this regime, the original global random field $\Gamma(\boldsymbol{x})$ is divided into $R$ local fields $\Gamma_r(\boldsymbol{x}),1\leq r\leq R$. Define the spatial correlation distance of the $r$-th subregion as $L_c^r$. Under the assumption of second-order stationarity, the covariance\footnote{Note that the covariance is defined with respect to the underlying continuous space before discretization, and the local correlation distance $L_c^r$ corresponds to the range value $a$ in the semivariogram defined in \eqref{Exponential}.}between two locations $\boldsymbol{x},\boldsymbol{x}'$ is
	
	\begin{align}
		C_{r}(\boldsymbol{\tau}) = \sigma^2_r \cdot e^{-\frac{\|\boldsymbol{\tau}\|}{L_c^r}}, \quad 1\leq r\leq R,  \label{local} 
	\end{align} where $\boldsymbol{\tau}=\boldsymbol{x}-\boldsymbol{x}'$ is the spatial vector from location $\boldsymbol{x}$ to $\boldsymbol{x}'$. Define the discretization granularity for the $r$-th subregion as $\Delta_r$, and the volume of the $r$-th subregion as $V_r$. Then, the number of grid points allocated for the $r$-th subregion is
	
	\begin{equation}
		n_r= |\mathcal{U}_r|=\frac{V_r}{\Delta_r^3}. \label{n_Delta}
	\end{equation}

	The power spectral density (PSD) of each subregion is the 3D-Fourier transform \cite{RandomField} of the covariance function \eqref{local}, i.e.,
	
	\begin{align}
		S_r (\boldsymbol{k})=\int_{\mathbb{R}^3} C_{r}(\boldsymbol{\tau}) e^{-i\boldsymbol{k}\cdot \boldsymbol{\tau}} d  \boldsymbol{\tau} 
		=\frac{8\pi \sigma^2_r {(L_c^r)}^3}{\left( 1+(2\pi L_c^r \| \boldsymbol{k}\|)^2 \right)^2}. \label{Spectrum}
	\end{align}

	The detailed derivation for $S_r (\boldsymbol{k})$ is given in the Appendix. The information-loss from the original continuous Gaussian field to its discrete grid representation arises from the truncation of the PSD, where the high-frequency components are lost. Define the cutoff frequency for the $r$-th subregion as ${k}_{\text{cut}}^r$. According to the Nyquist sampling theorem, when the discretization granularity $\Delta_r \leq \frac{1}{2k_{\text{cut}}^r}$, we can reconstruct a spatial field $\Gamma_r^{t}(\boldsymbol{x})$ whose PSD is 
	
	\begin{equation}
		S_r^t (\boldsymbol{k})  = 
		\begin{cases} 
			S_r (\boldsymbol{k}), &\|\boldsymbol{k}\|\leq k_{\text{cut}}^r, \\
			0 , &\|\boldsymbol{k}\|> k_{\text{cut}}^r,
		\end{cases}
	\end{equation}where the superscript $t$ represents ``truncated''. For ease of analysis, we constrain that
	
	\begin{equation}
		\Delta_r = \frac{1}{2k_{\text{cut}}^r}.   \label{K_Delta}
	\end{equation}
	
	The detailed reconstruction process for $\Gamma_r^{t}(\boldsymbol{x})$ is given in the Appendix. Define the error between $\Gamma_r(\boldsymbol{x})$ and $\Gamma_r^{t}(\boldsymbol{x})$ as

	\begin{equation}
		{\epsilon}_r({\boldsymbol{x}})=\Gamma_r(\boldsymbol{x}) - \Gamma_r^{t}(\boldsymbol{x}).
	\end{equation}
	
	According to the Wiener-Khinchin Theorem, the mean-squared value of a stationary process is equal to the integral of its PSD over the frequency domain. Therefore, we have
	
	\begin{align}
		&\mathbb{E}\left[|{\epsilon}_r({\boldsymbol{x}})|^2 \right] \notag      \\
		= & \int_{\boldsymbol{k}\in \mathbb{R}^3} \left[ S_r (\boldsymbol{k})-S_r^t (\boldsymbol{k}) \right]  d\boldsymbol{k}  
		\notag      \\
		=& \int_{\|\boldsymbol{k}\|\geq k_{\text{cut}}^r }   S_r (\boldsymbol{k}) d\boldsymbol{k}   \notag      \\
		=&\int_0^{2\pi}\int_0^{\pi}\int_{k_{\text{cut}}^r}^{\infty} \frac{8\pi \sigma^2_r(L_c^r)^3}{\left( 1+(2\pi (L_c^r)^3 \rho)^2 \right)^2}  \rho^2\sin{\phi} d\rho d\phi d\theta,
	\end{align} where $\rho=\|\boldsymbol{k}\|$ is the module of the spatial frequency $\boldsymbol{k}$. Given the assumption of second-order stationarity, we have $\mathbb{E}\left[|{\epsilon}_r({\boldsymbol{x}})|^2 \right]=\mathbb{E}\left[|{\epsilon}_r({\boldsymbol{x}'})|^2 \right]=D_r$ for $\forall \boldsymbol{x},\boldsymbol{x}' \in \mathcal{D}_r$. Define $t=2\pi L_c^r \rho$, and $t_r=2\pi L_c^r k_{\text{cut}}^r$. Then, we have

	\begin{align}
		D_r= &\int_0^{2\pi}\int_0^{\pi}\int_{t_r}^{\infty} \frac{ \sigma^2_r t^2}{ \pi^2 \left( 1+t^2 \right)^2}  \sin{\phi} dt d\phi d\theta \notag \\
		= & \frac{4\sigma^2_r}{\pi}\int_{t_r}^{\infty} \frac{ t^2 }{\left( 1+t^2 \right)^2}  dt \notag \\
		= &\frac{4\sigma^2_r}{\pi}\left (\int_{t_r}^{\infty}\frac{1 }{\left( 1+t^2 \right)}-\frac{1 }{{\left( 1+t^2 \right)}^2} dt \right )\notag \\
		= &\frac{4\sigma^2_r}{\pi} \left(\frac{1}{2}\text{arctan} t -\frac{t }{2(1+t^2)}+C \bigg|_{t_r}^{\infty}  \right) \notag \\
		=&\frac{4\sigma^2_r}{\pi}\left( \frac{\pi}{4} -\frac{1}{2} \text{arctan} {t_r} +\frac{t_r}{2(1+t_r^2)}\right) \notag \\
		=&\frac{4\sigma^2_r}{\pi} \left( \frac{1}{2} \text{arctan} \frac{1}{t_r}+\frac{t_r}{2(1+t_r^2)} \right)  \notag \\
		=&\frac{4\sigma^2_r}{\pi} \left(\frac{1}{2}\text{arctan} \frac{1}{2\pi L_c^r k_{\text{cut}}^r}+\frac{\pi L_c^r k_{\text{cut}}^r}{1+4\pi^2(L^r_c)^2 (k^r_{\text{cut}})^2} \right). \label{D(x)}
	\end{align}
	
	Note that when $k_{\text{cut}}^r\to 0$, we have $\Delta_r\to \infty$, and the discretization loss at $\boldsymbol{x}$ is $\lim_{k^r_{\text{cut}}\to 0} D_r= \sigma^2_r$, corresponding to the maximum uncertainty when interpolating $\Gamma_r(\boldsymbol{x})$. On the other hand, when $k^r_{\text{cut}}\to \infty$, we have $\Delta_r\to 0$, and the discretization-loss at $\boldsymbol{x}$ is $\lim_{k^r_{\text{cut}}\to \infty } D_r=0$, suggesting an accurate grid characterization incorporating all spatial frequency components. Using \eqref{K_Delta}, we can express $t_r$ in \eqref{D(x)} as $t_r=\frac{\pi L_c^r}{\Delta_r}$. When $L_c^r \gg \Delta_r$, i.e., $t_r\gg 1$, we have $\text{arctan} \frac{1}{t_r}\approx  \frac{1}{t_r}$, and $\frac{t_r}{2(1+t_r^2)}\approx \frac{1}{2t_r}$. Using the above approximations and \eqref{n_Delta}, we can reformulate \eqref{D(x)} in terms of $n_r$ as

	\begin{equation}
		D_r\approx\frac{4\sigma^2_r\Delta_r}{\pi^2L_c^r}=\frac{4\sigma^2_r}{\pi^2L_c^r} \left( \frac{V_r}{n_r}\right)^\frac{1}{3}. \label{Dr2}
	\end{equation}
	
	The structure of \eqref{Dr2} is intuitively satisfying. Specifically, $D_r$ is negatively correlated with $L_c^r$, suggesting that a decrease in local correlation distance requires a larger $n_r$, i.e., a smaller $\Delta_r$, to avoid discretization-loss. The negative exponent $-\frac{1}{3}$ of $n_r$ corresponds to the dimension of the physical space, in the sense that a linear decrease in $D_r$ requires a cubic increase in $n_r$. Using \eqref{Dr2}, we denote the discretization-loss in the $r$-th subregion as $V_r \cdot D_r$, and we wish to minimize the total discretization-loss in the $R$ subregions by optimizing the number of grid points allocated for each subregion, or equivalently, the discretization granularity \eqref{n_Delta} and the cutoff frequency \eqref{K_Delta} of each subregion. The optimization problem is formulated as 
	
	\begin{subequations}
		\begin{align}
			\textup{(P2):}\quad 
			\min_{\substack {n_r},1\leq r\leq R }\quad &  \sum_{r=1}^R  	\zeta_r  n_r^{-\frac{1}{3}} \\
			\text{s.t.} \quad & \sum_{r=1}^R n_r =U,  
		\end{align}
	\end{subequations} where $\zeta_r=\frac{4\sigma^2_rV_r^\frac{4}{3}}{\pi^2L_c^r}$. In order to deal with the equality constraint, we apply the Lagrangian multiplier and formulate
	
	\begin{align}
		\mathcal{L}(n_r,\lambda )= &\sum_{r=1}^R  	\zeta_r  n_r^{-\frac{1}{3}} +\lambda \left(\sum_{r=1}^R n_r -U\right) \notag \\
		= &\sum_{r=1}^R \left(	\zeta_r  n_r^{-\frac{1}{3}} +\lambda n_r\right)-\lambda U. \label{lagrangian_}
	\end{align}
	
	By setting the partial derivatives of \eqref{lagrangian_} with respect to each $n_r$ and the multiplier $\lambda$ to zero, we have
	
	\begin{equation}
		\begin{cases}
			\frac{\partial\mathcal{L}(n_r,\lambda )}{\partial n_r} = -\frac{1}{3} \zeta_r  n_r^{-\frac{4}{3}} +\lambda=0 ,1\leq r\leq R \\
			\frac{\partial\mathcal{L}(n_r,\lambda )}{\partial \lambda} =\sum_{r=1}^R n_r =U.
		\end{cases}   
	\end{equation}  
	
	Solving the above equations, we obtain
	
	\begin{equation}
		\begin{cases}
			\lambda = U^{-\frac{4}{3}}\cdot \left(\sum_{r=1}^R \left( \frac{\zeta_r}{3} \right)^{\frac{3}{4}}  \right)^{\frac{4}{3}} ,\\
			n_r= U\cdot  \frac{ V_r\cdot  \left( \frac{\sigma_r^2}{L_c^r} \right)^{\frac{3}{4}}   }{ \sum_{r=1}^R  V_r\cdot  \left( \frac{\sigma_r^2}{L_c^r} \right)^{\frac{3}{4}} }. 
		\end{cases}    \label{allocation}
	\end{equation}

    Note that the fractions $\{n_r\}$ in \eqref{allocation} are ultimately relaxed into integers for the practical allocation of $U$. In summary, the proposed adaptive discretization minimizes the discretization-loss in the mean-squared sense, delineating the optimal structure of the underlying solution space for the optimization algorithms.

	\section{Simulation Results}
	
\begin{figure}[H]	
		\centering	
		\includegraphics[width=8cm,height=6cm]{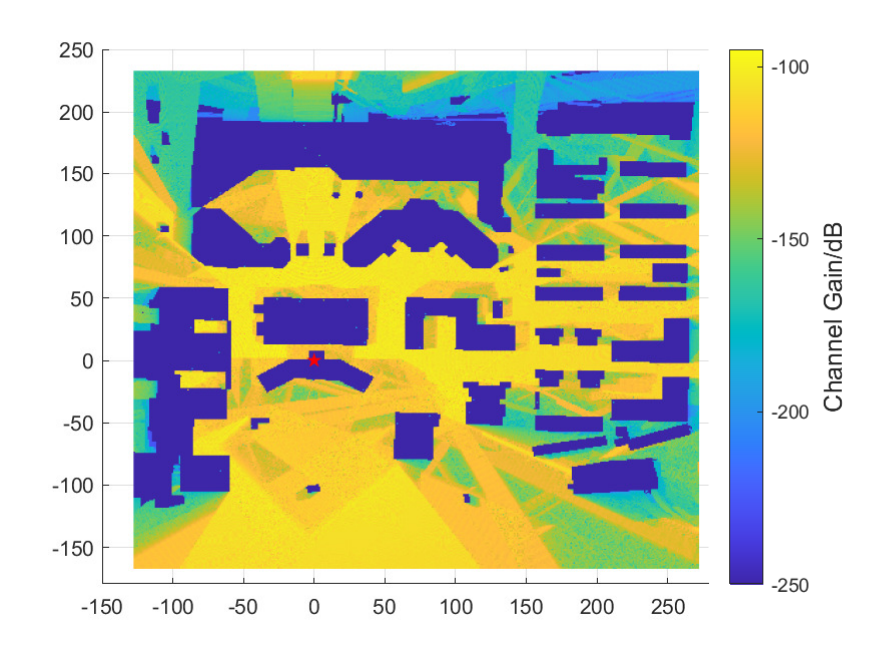}	
		\caption{The ground-truth mean function $\mu(\boldsymbol{x})$.}
		\label{fig:GT_CKM}
	\end{figure}

	As shown in Fig. \ref{fig:GT_CKM}, the ground-truth $\mu(\boldsymbol{x})$ is generated by the commercial ray-tracing software Wireless Insite. The reduced candidate set $\boldsymbol{\mathcal{U}}$ is constructed either by uniform sampling of the universal set $\boldsymbol{\mathcal{D}}$ or by using the proposed adaptive discretization strategy. The zero-mean $S(\boldsymbol{x})$ is synthesized via Cholesky decomposition of the covariance matrix analytically constructed based on \eqref{local}.

\subsection{Spatial Partitioning and Adaptive Allocation of Grid Points}

The proper partitioning of the physical space is crucial to the validity of the second-order stationarity assumption for ordinary Kriging. In this subsection, we partition the physical space into $R$ homogeneous subregions, and derive the mean value at each spatial location by exploiting the basic law of electromagnetic wave propagation. The standing assumption is that the mean value at each location follows a specific propagation (path-loss) model, and we can fit the parameters of the underlying path-loss model using the mean value at measured locations to predict that of unmeasured locations. Due to heterogeneous shadowing effects, the path-loss models vary across different subregions, and we employ the method of K-means \cite{xu2024much} to fit the path-loss model for each subregion.

By placing the BS at the origin, we denote the Euclidean distance from each location $\boldsymbol{x}_i$ to the BS as $\|\boldsymbol{x}_i\|$. Define $d_{i}=10\log_{10} \|\boldsymbol{x}_i\|$. Then we map each $\boldsymbol{x}_i$ to a coordinate value $(d_{i},\mu(\boldsymbol{x}_i))$. Due to the exponential power attenuation of electromagnetic waves, the mean channel gain $\mu(\boldsymbol{x})$ at a specific location is linearly correlated with the logarithmic value of the propagation distance to the BS, i.e., $\mu=\alpha d +\beta$. Therefore, we initialize $R$ lines in the form

	\begin{equation}
		\mu=\alpha_r d +\beta_r, 1\leq r \leq R.  \label{mapping1}
	\end{equation}

Each fitted line corresponds to the center of a cluster $C_r,1\leq r\leq R$, and we employ the K-means to associate each $(d_{i},\mu(\boldsymbol{x}_i))$ with a specific cluster $C_r$. As an expectation-maximization (EM) algorithm, the implementation of K-means can be divided into the E-step and the M-step. In the E-step, the distance between each measured point and each line is calculated, and each measured point is associated with its nearest cluster. Then, in the M-step, the least-square fitting is performed for each $\boldsymbol{\theta}_r =[\alpha_r,\beta_r]$, i.e.,
	
	\begin{equation}
		\boldsymbol{\theta}_r= \left( \boldsymbol{H}_r^T \boldsymbol{H}_r \right)^{-1}\boldsymbol{H}_r^T \boldsymbol{\mu}_r,
	\end{equation} where $	\boldsymbol{H}_r = \begin{bmatrix}
		-{d}_{1} & 1 \\
		\vdots & \vdots \\
		-{d}_{|C_r|} & 1
	\end{bmatrix}$, and $\boldsymbol{\mu}_r=[\mu(\boldsymbol{x}_1),\mu(\boldsymbol{x}_2),...,\mu(\boldsymbol{x}_{|C_r|})]$. After acquiring the $R$ clusters based on measurement data, each unmeasured location can be subsequently assigned to one of the $R$ clusters. For example, the cluster category of each unmeasured location can be determined by majority voting based on its $m$ nearest measured locations. Then, the mean channel gain at each unmeasured location is estimated based on the fitted lines in \eqref{mapping1}. 
	

Each cluster $C_r$ essentially corresponds to a specific path-loss model, and locations with similar path-loss exponents and homogeneous shadowing effects are grouped into the same subregion by the method of K-means. Within each cluster, the channel gain exhibits stable spatial covariance structure, thereby validating the second-order stationarity assumption. As revealed in \eqref{allocation}, each subregion essentially competes for grid resources with their respective $ V_r\cdot  \left( \frac{\sigma_r^2}{L_c^r} \right)^{\frac{3}{4}}$. Define $g(V_r,L_c^r)= V_r\cdot  \left( \frac{\sigma_r^2}{L_c^r} \right)^{\frac{3}{4}}$. Then, we have, $\frac{\partial g(V_r,L_c^r)}{\partial V_r}= \sigma_r^{\frac{3}{2}}\left(L_c^r\right)^{-\frac{3}{4}}$, and $\frac{\partial g(V_r,L_c^r)}{\partial L_c^r}=-\frac{3}{4}V_r\sigma_r^{\frac{3}{2}}\left(L_c^r\right)^{-\frac{7}{4}}$. Under typical parameter settings, i.e., $4000 \leq V_r \leq 10000$ and $5\leq L_c^r \leq 25$, the maximum $\left|\frac{\partial g(V_r,L_c^r)}{\partial V_r}\right|$ is on the scale of $0.3\sigma_r^{\frac{3}{2}}$. On the other hand, the smallest $\left|\frac{\partial g(V_r,L_c^r)}{\partial L_c^r}\right|\approx 10\sigma_r^{\frac{3}{2}}$, suggesting that the decrease in $L_c^r$ dominates the increase in $g(V_r,L_c^r)$, and therefore regions with short $L_c^r$, such as region 1 in Table. \ref{tab:Subregion Parameters}, are expected to gain a larger share of $U$.

	\begin{table}[t]
		\centering
		\caption{Subregion Parameters}
		\label{tab:Subregion Parameters}
		\begin{tabular}{l|cccc}
			\hline
			& $V_r$($\text{m}^3$) & $L_c^r$($\text{m}$) & $n_r$ &  $\Delta_r$ \\
			\hline  
			1    & 4749  & 6.5&  494  &  3  \\
			2   & 10242  & 22.1& 425  &  5  \\
			3    &  8586 &17.5 & 424 &   4  \\
			4    & 10053  & 21.7& 422  & 5   \\
			5    &  8415  & 16.7& 432  & 5 \\
			6    & 8322  & 14.0&  486 &  4 \\
			7    & 7632  & 17.6&  375 &  5  \\
			8   &   7152  & 14.6& 405 &  4 \\
			9    &  11169 & 25.0& 422  & 6 \\
			10    & 5502  & 10.1& 411  & 3 \\
			\hline
		\end{tabular}
	\end{table}

    \subsection{The Influence of Semivariogram Type}
	
In this subsection, we analyze the influence of semivariogram type on the measurement pattern, and compare their performance on the reduction of global AMSE using the SA. Aside from the exponential semivariogram introduced in previous sections, we consider two other semivariograms, namely the Gaussian semivariogram $\gamma_{\text{Gau}}(h) = C_0 + C\left(1 - e^{-\frac{h^2}{a^2}}\right)$, and the spherical semivariogram $\gamma_{\text{Sph}}(h) = 
\begin{cases}
C_0 + C\left( \frac{3h}{2a} - \frac{h^3}{2a^3} \right), & 0 \le h \le a, \\
C_0 + C, & h > a.
\end{cases}$

\begin{figure}[H]
		\centering  
		\subfigbottomskip=2pt 
		\subfigcapskip=-5pt 
		\subfigure[Greedy search with the exponential semivariogram, $N$=200.]{
			\includegraphics[width=4cm,height=3cm]{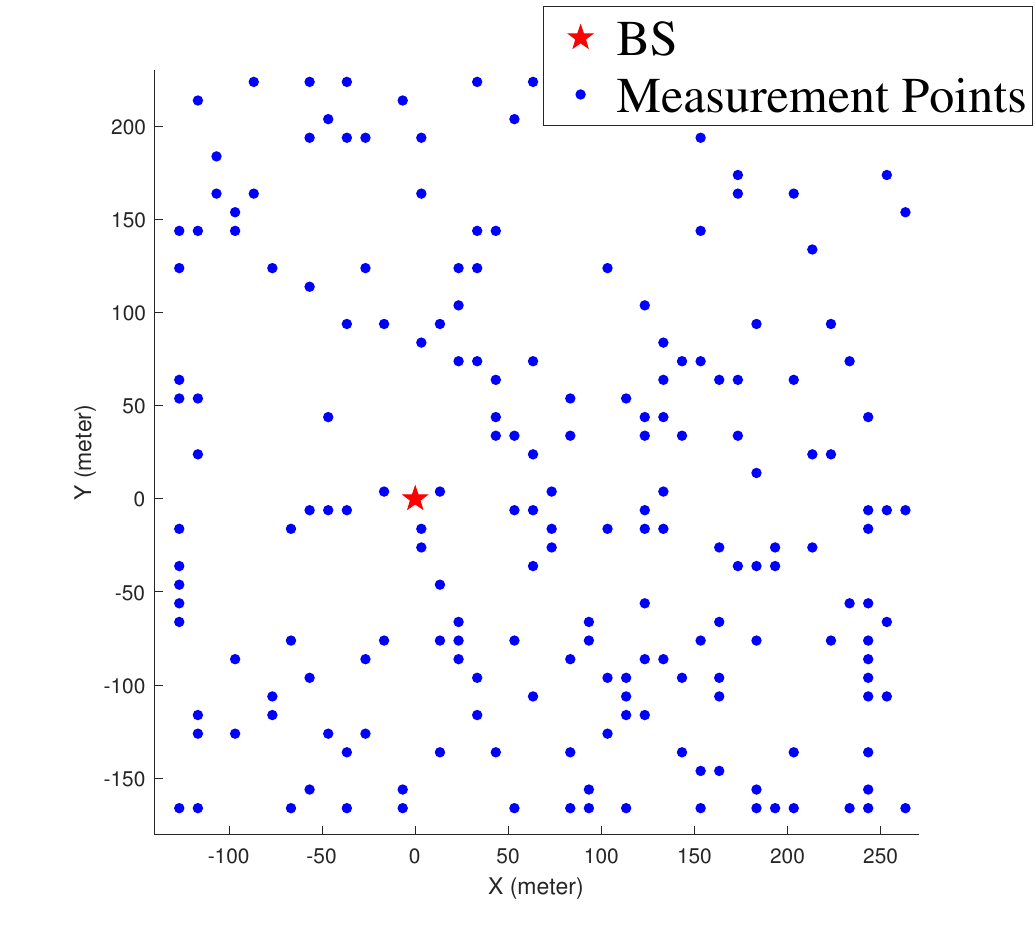}}     
		\subfigure[SA search with the exponential semivariogram, $N$=200.]{
			\includegraphics[width=4cm,height=3cm]{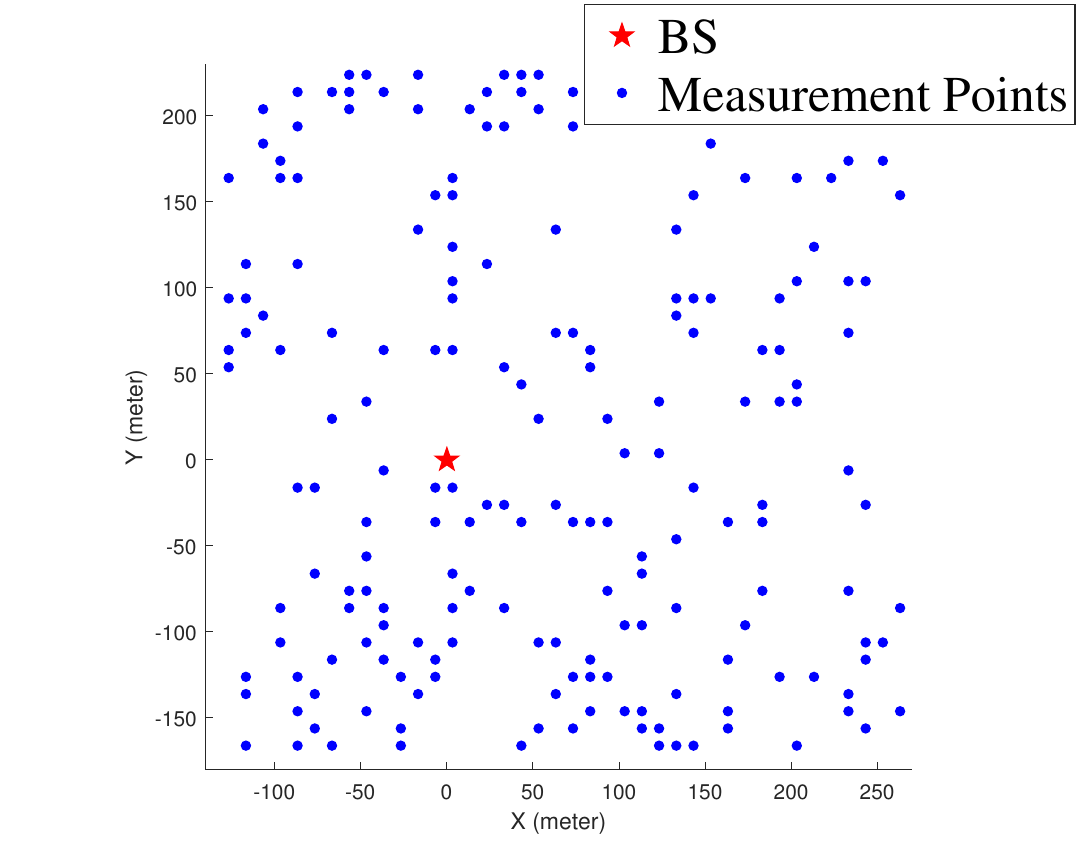}}
		\subfigure[SA search with the Gaussian semivariogram, $N$=200.]{
			\includegraphics[width=4cm,height=3cm]{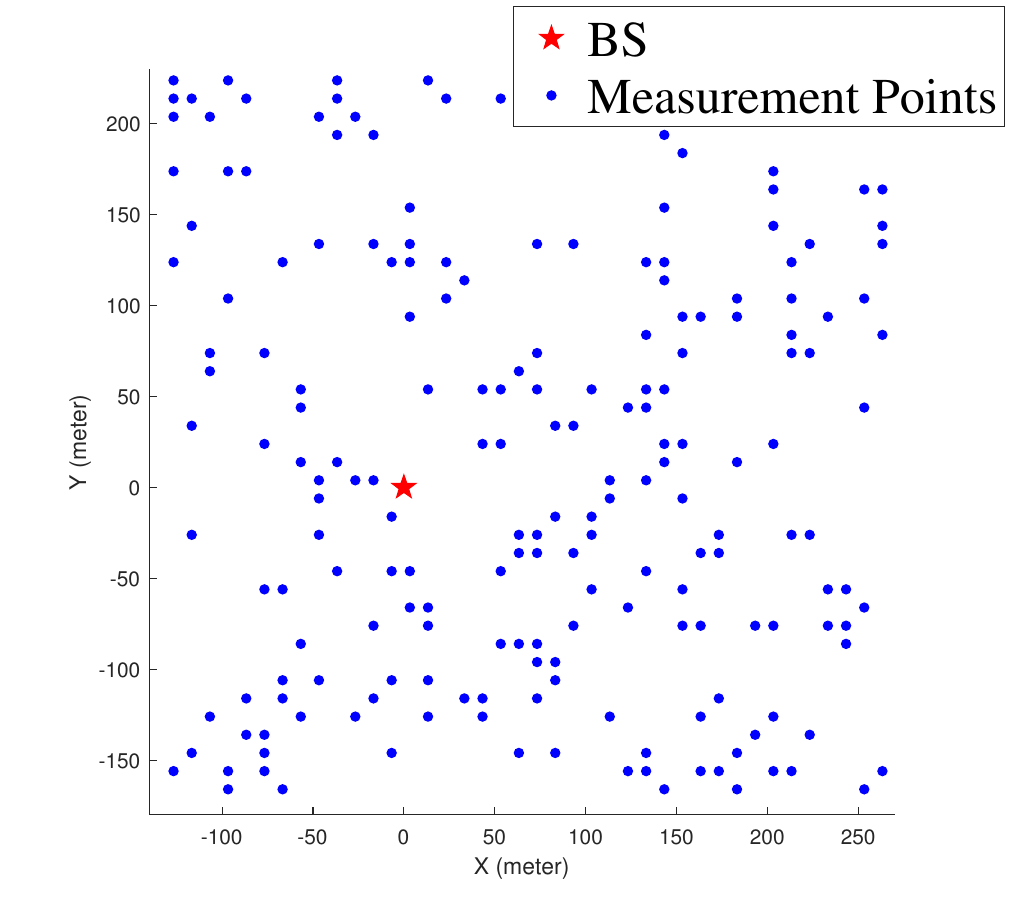}}
		\subfigure[SA search with the spherical semivariogram, $N$=200.]{
			\includegraphics[width=4cm,height=3cm]{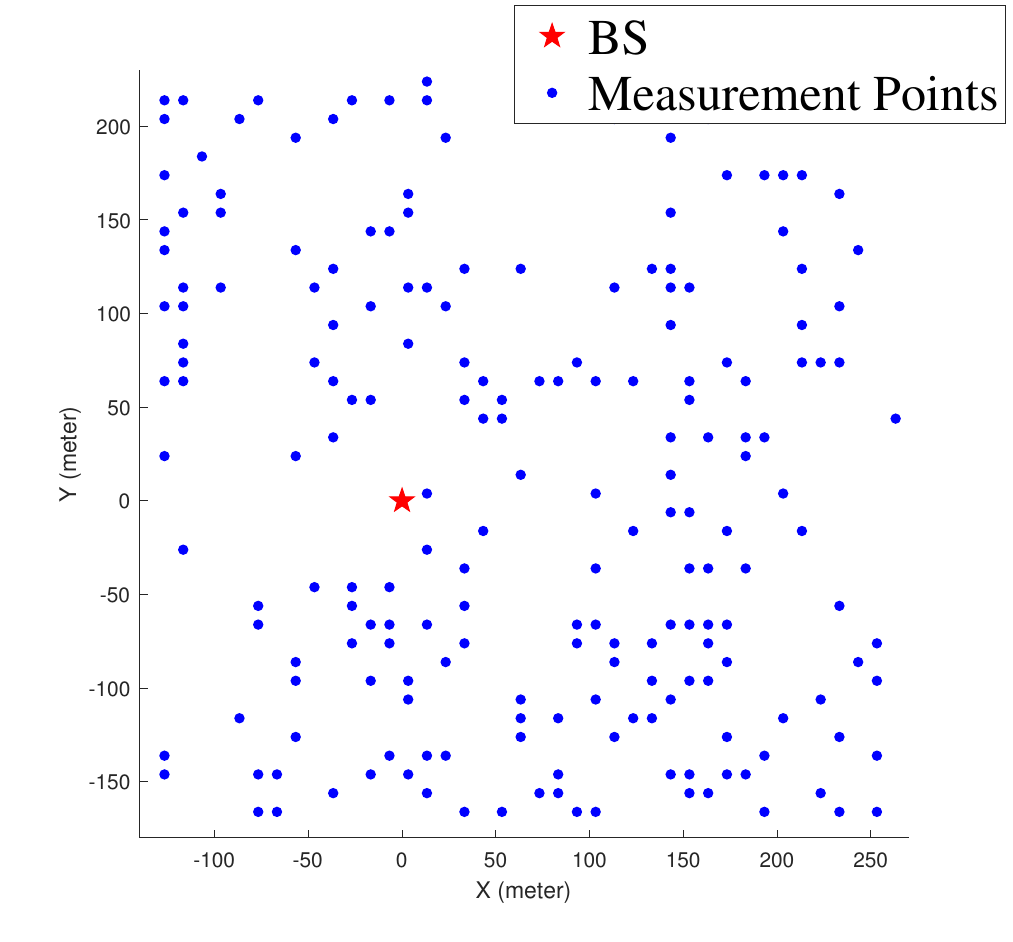}}
		\caption{Visualization of the measurement set $\boldsymbol{\mathcal{S}}$.} \label{fig:PPP}
	\end{figure}

As shown in Fig. \ref{fig:PPP}, the number of local clusters in the measurement pattern of the exponential model is much larger than that in the other two models. This phenomenon can be explained by the mathematical structure of the three semivariograms. Specifically, we perform the Taylor expansion as $h \to 0$, and obtain $\gamma_{\text{Exp}}(h) \approx \frac{3h}{a}-\frac{9h^2}{2a^2} + O(h^3)$, where the derivative at $h=0$ is $\frac{3}{a}$; 
$\gamma_{\text{Gau}}(h) \approx \left(\frac{3h}{a}\right)^{\!2} - \frac{1}{2}\left(\frac{3h}{a}\right)^{\!4} + O(h^6)$, where the derivative at $h=0$ is $0$; 
$\gamma_{\text{Sph}}(h) \approx \frac{3h}{2a} - \frac{1}{2}\left(\frac{h}{a}\right)^{\!3} + O(h^5)$, where the derivative at $h=0$ is $\frac{3}{2a}$. When the lag value $h$ is small (e.g., $h < 0.5a$), the derivative value of the exponential semivariogram is larger compared to the other two models. Therefore, once the semivariogram $\gamma_{\text{Exp}}(h)$ exceeds a threshold, the exponential model can readily identify this high-variance location even if the lag value $h$ is small, thus producing the local measurement clusters. In terms of the Gaussian model, the Taylor expansion above reveals that the it is exceedingly ``flat'' around the origin, suggesting that the channel gains of neighboring locations at small $h$ are highly correlated. Consequently, adjacent measurement locations tend to ``repel'' each other to avoid local densification, resulting in a global uniform distribution. For the spherical model, the spatial correlation drops directly to zero outside the range value $a$, i.e., for each measurement point, its effective influence area is characterized by a sphere centered on the measurement point with a radius of $a$. The unmeasured locations at the joint boundary of neighboring influence spheres suffer from considerably larger estimation variances, and therefore would be preferentially selected by the optimization algorithms. Consequently, the determined $\boldsymbol{\mathcal{S}}$ exhibits a sphere-packing distribution pattern in which the influence spheres of different measurement points are arranged with minimal overlaps.

\begin{figure}[H]	
		\centering	
		\includegraphics[width=8.5cm,height=6cm]{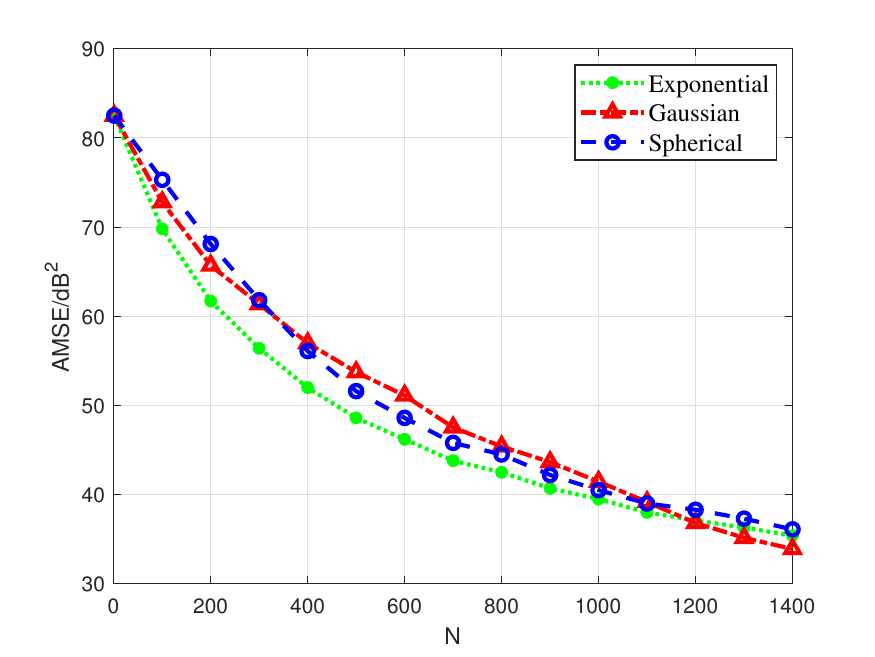}	
		\caption{Comparison of different semivariograms.}
		\label{fig:Semivariogram}
	      \end{figure}

In terms of the performance on AMSE-reduction shown in Fig. \ref{fig:Semivariogram}, the curves exhibit a distinct three-stage decreasing trend. In the first stage, i.e., $N<100$, the average distance between adjacent measurement points could far exceed the range value $a$. Due to the ``cutoff'' of correlation at $h=a$, the spherical semivariogram significantly undermines the influence area of each measurement point, and shows the worst performance on AMSE-reduction. In the second stage, as $N$ increases, both the exponential and the spherical semivariogram start to target high-variance subregions by forming local measurement clusters, contributing to a steady decline in global AMSE. The Gaussian semivariogram, on the other hand, precludes such local densification, and yields the lowest AMSE-reduction. In the third stage, i.e., $N>1000$, the advantage of the Gaussian semivariogram starts to manifest. Thanks to the repulsive effect mentioned above, the measurement locations determined by the Gaussian model exhibit the most uniform distribution, enabling the exploration of measurement holes at the border of the considered space. Consequently, the AMSE-reduction of the Gaussian model exceeds that of the other two models in this stage.

	\subsection{AMSE Analysis}

	In this subsection, we discuss different combinations of discretization strategies and optimization algorithms. The semivariogram in this subsection is fixed as the exponential type to comply with the exponential covariance function, and the AMSE is calculated with respect to the universal set $\boldsymbol{\mathcal{D}}$.

\begin{figure}[H]	
		\centering	
		\includegraphics[width=8.5cm,height=6cm]{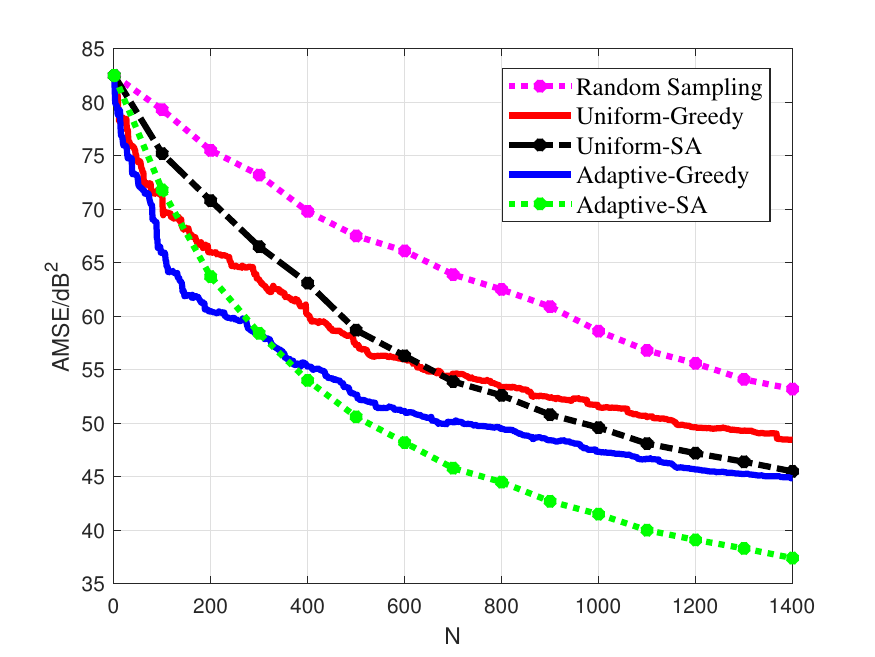}	
		\caption{Average MSE of ordinary Kriging-based optimization combinations.}
		\label{fig:AMSE}
	      \end{figure}

    As shown in Fig. \ref{fig:AMSE}, when the size of $\boldsymbol{\mathcal{S}}$ is small, the greedy algorithm precisely identifies unmeasured locations with the highest estimation variance, thus achieving a faster AMSE-reduction than the SA. However, as $N$ increases, it falls into the trap of local optima, where newly selected unmeasured locations excessively cluster around existing local high-variance regions, resulting in a progressively smaller AMSE-reduction. On the other hand, thanks to the global exploration capability of the Metropolis criterion, SA successfully avoids local optima and outperforms the greedy algorithm after a threshold. Lastly, random sampling generates the measurement set $\boldsymbol{\mathcal{S}}$ in a completely stochastic manner without utilizing any prior spatial information, potentially causing redundant sampling in low-AMSE regions and sparse coverage across high-AMSE measurement holes, resulting in the smallest AMSE-reduction. Fig. \ref{fig:AMSE} also demonstrates the superiority of the proposed adaptive discretization strategy. In uniform discretization, subregions with large correlation distances are allocated an excessive number of grid points, and measurement at these low-value locations provides a low AMSE-reduction. Meanwhile, the insufficient number of grid points allocated in subregions with small correlation distances may fail to resolve the variance-hotspots, precluding the optimization algorithms from exploiting locations contributing to global high AMSE-reduction. Compared to uniform discretization, adaptive discretization precisely directs the limited grid resource to regions of strong spatial variability, promoting the identification of locations that potentially yield the maximum AMSE‑reduction.
	
	\begin{figure}[H]	
		\centering	
			\includegraphics[width=8.5cm,height=6cm]{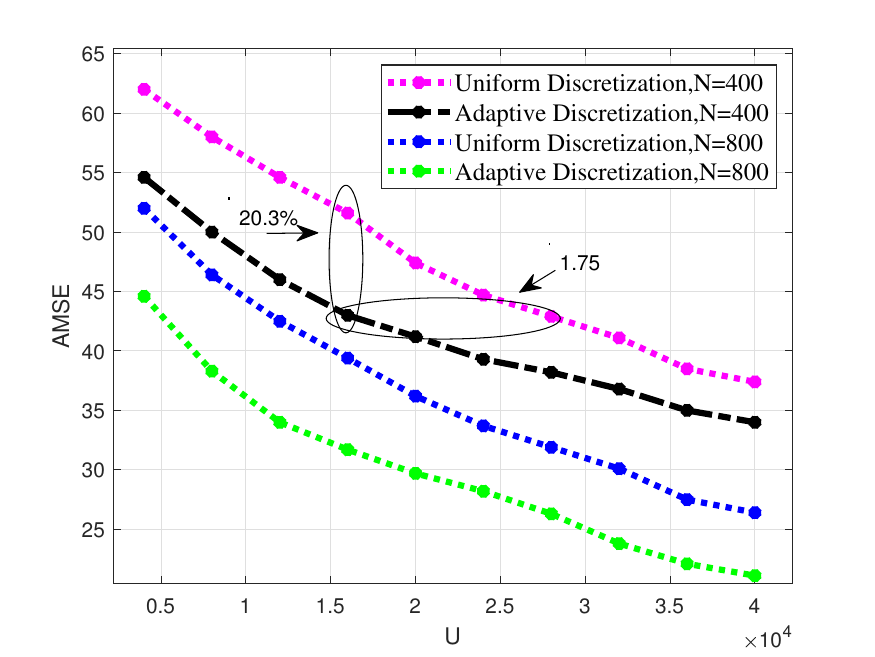}	
		\caption{Average MSE under different universal set size $U$.}
		\label{fig:Complexity}
	\end{figure}

	Next, we demonstrate the performance superiority of the proposed adaptive discretization strategy under different candidate set size $U$. As illustrated in Fig. \ref{fig:Complexity}, adaptive discretization strategy provides a consistent performance gain in AMSE-reduction under the same universal set size $U$. Specifically, when $U=16000,N=400$, the AMSE of uniform discretization is $20\%$ higher than that of adaptive discretization. Equivalently, one can conclude that under the condition of equal AMSE-reduction, adaptive discretization is superior in terms of computational complexity. Specifically, to achieve a global AMSE of 43, the required candidate set size $U$ for uniform discretization is 1.75 times that of adaptive discretization, corresponding to a 5.36-fold increase in computational complexity.

\section{Conclusions}
	
	In this paper, we studied the design of spatial measurements for CGM construction. Specifically, we decomposed the problem into two hierarchical components, namely spatial discretization and combinatorial optimization, aiming at selecting a subset of locations for channel measurements, so as to minimize the AMSE of the global CGM construction. Aside from the traditional uniform discretization, we developed an adaptive discretization strategy based on the Gaussian random field theory, and further proposed two optimization algorithms for solving the formulated combinatorial optimization problem. By adopting different combinations of discretization strategies and optimization algorithms, we validated the important role of spatial discretization on the efficiency of global AMSE-reduction, providing useful guidance for practical spatial measurements for CGM construction.

	\appendix

	For ease of exposition, we denote $C_{r}(\boldsymbol{\tau})$ by $C(\boldsymbol{\tau})$, and $S_r(\boldsymbol{k})$ by $S(\boldsymbol{k})$. Given that $C(\boldsymbol{\tau})$ is a radial function, in the sense that the value of $C(\boldsymbol{\tau})$ depends only on $\|\boldsymbol{\tau}\|$, we apply the radial symmetry property of the 3D Fourier transform, and define $r=\|\boldsymbol{\tau}\|$, $k=\|\boldsymbol{k}\|$. We use a spherical coordinate system and align the direction of $\boldsymbol{k}$ with the $z$-axis to yield $ \boldsymbol{k}\cdot \boldsymbol{\tau}=kr\cos{\phi}$. Then, we have

	\begin{align}
		S(\boldsymbol{k})=&\int_0^{2\pi}\int_0^{\pi}\int_{0}^{\infty} C(r) e^{-ikr\cos{\phi}} r^2 \sin{\phi} dr d\phi d \theta \notag \\
		=& 2\pi \int_{0}^{\infty} C(r)r^2 \left(\int_0^{\pi} e^{-ikr\cos{\phi}} \sin{\phi}d\phi\right) dr. \label{Appendix1}
	\end{align}
	
	Let $u=\cos{\phi}$, we have 
	
	\begin{align}
		\int_0^{\pi} e^{-ikr\cos{\phi}} \sin{\phi}d\phi =& \int_1^{-1} e^{-ikru}(-du) \notag \\  
		=& \int_{-1}^{1} e^{-ikru}du \notag \\  
		=&\frac{2\sin({kr})}{kr}, 
	\end{align} and \eqref{Appendix1} can be simplified as
	
	\begin{align}
		S(\boldsymbol{k})=& \frac{4\pi}{k}   \int_{0}^{\infty}  C(r) \sin({kr})r dr \notag \\
		=&\frac{4\pi\sigma^2}{k}\int_{0}^{\infty} e^{-\frac{r}{L_c}}\sin({kr})r dr. 
	\end{align} 
	
	Then we use the integral equation 
	
	\begin{equation}
		\int_{0}^{\infty} e^{-\alpha r}\sin({\beta r})r dr = \frac{2\alpha \beta}{(\alpha^2+\beta^2)^2},
	\end{equation} to yield the PSD \eqref{Spectrum} of the Gaussian field $\Gamma_r(\boldsymbol{x})$ defined in the $r$-th subregion. As defined in Section VI, $\Gamma_{r}^t(\boldsymbol{x})$ is the truncated field with the PSD
	
	\begin{equation}
		S_r^t (\boldsymbol{k})  = 
		\begin{cases} 
			S_r (\boldsymbol{k}), &\|\boldsymbol{k}\|\leq k_{\text{cut}}^r, \\
			0 , &\|\boldsymbol{k}\|> k_{\text{cut}}^r.
		\end{cases}
	\end{equation}
	
	Define the spatial periodic sampling function as
	
	\begin{equation}
		s_r(\boldsymbol{x})=\sum_{\boldsymbol{n}\in\mathbb{Z}^3}  \delta( \boldsymbol{x}-\boldsymbol{n}\Delta_r), 
	\end{equation} where $\delta(\boldsymbol{x})$ is the spatial Dirac function. The sampled field is denoted as
	
	\begin{equation}
	\Gamma_{r}^s(\boldsymbol{x})  = \Gamma_{r}^t(\boldsymbol{x}) \cdot    s_r(\boldsymbol{x})=\Gamma_{r}^t(\boldsymbol{n}\Delta_r)\delta( \boldsymbol{x}-\boldsymbol{n}\Delta_r).     
	\end{equation}

	According to the frequency-convolution theorem of the Fourier transform, the PSD of $\Gamma_{r}^s(\boldsymbol{x})$ is 
	
	\begin{align}
		S_r^s (\boldsymbol{k}) =&  S_r^t (\boldsymbol{k}) * \mathcal{F}\left[ s_r(\boldsymbol{x}) \right]  \notag \\
		=& \frac{1}{\Delta_r^3} \sum_{\boldsymbol{n}\in\mathbb{Z}^3}  S_r^t \left(\boldsymbol{k}-\frac{\boldsymbol{n}}{\Delta_r}\right). 
	\end{align}
	
	As long as $2k^r_{\text{cut}} \leq  \frac{1}{\Delta_r}$, $\Gamma_{r}^t(\boldsymbol{x})$ can be fully reconstructed from the interpolation of $\Gamma_{r}^s(\boldsymbol{x})$, i.e., 
	
	\begin{align}
		\Gamma_{r}^t(\boldsymbol{x})=\sum_{\boldsymbol{n}\in\mathbb{Z}^3}\Gamma_{r}^t(\boldsymbol{n}\Delta_r)\text{sinc}_3 \left(\frac{\boldsymbol{x}-\boldsymbol{n}\Delta_r}{\Delta_r}\right),
	\end{align} where $\text{sinc}_3(\boldsymbol{x})= \prod_{i=1}^3\text{sinc}(x_i)$.

	\bibliographystyle{IEEEtran}  
	\bibliography{CKM_Temporal_Spatial}

\end{document}